\documentclass[lettersize,journal]{IEEEtran}
\usepackage{cite}
\usepackage{amsmath,amssymb,amsfonts}
\usepackage{dsfont}
\usepackage{algorithm}
\usepackage{bm}
\usepackage{algorithmic}
\usepackage{graphicx}
\usepackage[caption=false]{subfig}
\usepackage{textcomp}
\usepackage{xcolor}
\usepackage{float}
\usepackage{tabularx}
\usepackage{xurl}
\usepackage{nomencl}
\makenomenclature
\newcommand\ph{$\phantom{1}$} 
\allowdisplaybreaks
\usepackage{lipsum}
\usepackage[hidelinks]{hyperref}
\def\BibTeX{{\rm B\kern-.05em{\sc i\kern-.025em b}\kern-.08em
    T\kern-.1667em\lower.7ex\hbox{E}\kern-.125emX}}
\begin{document}

\title{
Economic MPC with an Online Reference Trajectory for Battery Scheduling Considering Demand Charge Management
{\footnotesize \textsuperscript{}}
}
\author{Cristian Cortes-Aguirre$^*$,~\IEEEmembership{Graduate Student Member,~IEEE}, Yi-An Chen$^*$, Avik Ghosh,~\IEEEmembership{Graduate Student Member,~IEEE}, Jan Kleissl, and Adil Khurram,~\IEEEmembership{Member,~IEEE}
\thanks{$^*$: These authors contributed to the paper equally}
}

\maketitle

\begin{abstract}
Monthly demand charges form a significant portion of the electric bill for microgrids with variable renewable energy generation. A battery energy storage system (BESS) is commonly used to manage these demand charges. Economic model predictive control (EMPC) with a reference trajectory can be used to dispatch the BESS to optimize the microgrid operating cost. Since demand charges are incurred monthly, EMPC requires a full-month reference trajectory for asymptotic stability guarantees that result in optimal operating costs. However, a full-month reference trajectory is unrealistic from a renewable generation forecast perspective. Therefore, to construct a practical EMPC with a reference trajectory, an EMPC formulation considering both non-coincident demand and on-peak demand charges is designed in this work for $24$ to $48$~\textcolor{black}{h} prediction horizons. The corresponding reference trajectory is computed at each EMPC step by solving an optimal control problem over $24$ to $48$~\textcolor{black}{h} reference (trajectory) horizon. Furthermore, BESS state of charge regulation constraints are incorporated to guarantee the BESS energy level in the long term. Multiple reference and prediction horizon lengths are compared for both shrinking and rolling horizons with real-world data. The proposed EMPC with $48$~hour rolling reference and prediction horizons outperforms the traditional EMPC benchmark with a $2$\% reduction in the annual cost, proving its economic benefits.

\end{abstract}


\nomenclature{$\mathbb{T}$}{Set of discrete time points}
\nomenclature{$t$}{Time point}
\nomenclature{$\mathcal{T}_t$}{Set of all time points in the month that $t$ belongs to}
\nomenclature{$\mathcal{T}_{\text{NC},t}$, $\mathcal{T}_{\text{OP},t}$}{Set of all time points corresponding to NC/OP demand charge periods for the month $\mathcal{T}_t$}
\nomenclature{$\mathcal{T}_{\text{MPC},t}$, $\mathcal{T}_{\text{R},t}$}{Set of all time points corresponding to the prediction/reference horizon starting at the time point $t$}
\nomenclature{$\tau_{\text{NC},t}$, $\tau_{\text{OP},t}$}{Final time point of the set $\mathcal{T}_{\text{NC},t}$/$\mathcal{T}_{\text{OP},t}$}
\nomenclature{$\tau_{\text{MPC},t}$, $\tau_{\text{R},t}$}{Final time point of the set $\mathcal{T}_{\text{MPC},t}$/$\mathcal{T}_{\text{R},t}$}
\nomenclature{$\tau_{e,t}$}{Final time point of a specific (future) horizon}
\nomenclature{$\mathcal{C}_1$}{Energy cost}
\nomenclature{$u_1$}{Demand}
\nomenclature{$u_2$}{BESS charging/discharging power}
\nomenclature{$\eta$}{Roundtrip efficiency}
\nomenclature{$R_{\text{EC}}$}{Energy charge rate}
\nomenclature{$\mathcal{C}_2$}{Demand charge cost}
\nomenclature{$\hat{P}_{\text{NC}}$, $\hat{P}_{\text{OP}}$}{Monthly DCT for NC/OP demand charge}
\nomenclature{$P_{\text{NC}}$, $P_{\text{OP}}$}{NC/OP peak demand within the set $\{t,\dots,\tau_{e,t}\}$}
\nomenclature{$R_{\text{NC}}$, $R_{\text{OP}}$}{NC/OP demand charge rate}
\nomenclature{$\sigma_{\text{NC}}$, $\sigma_{\text{OP}}$}{Binary variable to identify the beginning of $\mathcal{T}_{\text{NC},t}$/$\mathcal{T}_{\text{OP},t}$}
\nomenclature{$\mathbb{T}_{\rm NC}$, $\mathbb{T}_{\rm OP}$}{Set of the first time points of all $\mathcal{T}_{\text{NC},t}$/$\mathcal{T}_{\text{OP},t}$}
\nomenclature{$x$}{BESS SOC}
\nomenclature{$\Delta T$}{Time step}
\nomenclature{$\bm{u_1}$}{Vector of $u_1(t)$ for all $t \in \mathbb{T}$}
\nomenclature{$L_{f}$, $\text{PV}_{f}$}{Forecasted load/PV generation}
\nomenclature{$\overline{u}_{2}$}{BESS power capacity}
\nomenclature{$\overline{x}$, $\underline{x}$}{BESS SOC limits}
\nomenclature{$\text{BESS}_{en}$}{BESS energy capacity}
\nomenclature{$\hat{\tau}_{\text{MPC},t}$, $\hat{\tau}_{\text{R},t}$}{Time point where the $50$\% SOC low threshold is placed in a prediction/reference horizon}
\nomenclature{$y_{\rm NC}$, $y_{\rm OP}$}{NC/OP peak demand tracker}
\nomenclature{$z$}{Augmented state of the proposed EMPC formulation}
\nomenclature{$L$}{Stage cost}
\nomenclature{$z_r$}{State and control inputs used as a reference trajectory}
\nomenclature{$x_r$}{Reference trajectory BESS SOC}
\nomenclature{$u_{r1}$}{Reference trajectory's demand}
\nomenclature{$u_{r2}$}{Reference trajectory's BESS charging/discharging power}
\nomenclature{$k$, $k'$}{Prediction/Reference horizon time point}
\nomenclature{$\hat{k}$}{Time point where the reference trajectory was computed}
\nomenclature{$\hat{y}_{\text{NC}}$, $\hat{y}_{\text{OP}}$}{NC/OP reference trajectory peak}
\nomenclature{$\check{y}_{\text{NC}}$, $\check{y}_{\text{OP}}$}{Future NC/OP reference trajectory peak}
\nomenclature{$V_f$}{Terminal cost function}
\nomenclature{$\bm{u_{r1}}$}{Vector of $u_{r1}(t)$ for all $t \in \mathbb{T}$}
\nomenclature{$T_{\text{R}}$, $T_{\text{MPC}}$}{Reference/Prediction horizon length}
\nomenclature{MG}{Microgrid}
\nomenclature{SOC}{State of charge}
\nomenclature{VRE}{Variable renewable energy}
\nomenclature{EV}{Electric vehicle}
\nomenclature{PV}{Photovoltaic}
\nomenclature{NC}{Non coincident}
\nomenclature{OP}{On peak}
\nomenclature{MPC}{Model predictive control}
\nomenclature{EMPC}{Economic MPC}
\nomenclature{DCT}{Demand charge threshold}
\nomenclature{DCM}{Demand charge management}
\nomenclature{BESS}{Battery energy storage system}
\nomenclature{SAM}{System advisor model}
\nomenclature{NT, WT}{Without/With peak demand tracking}
\nomenclature{NCDC}{NC demand charge cost}
\nomenclature{OPDC}{OP demand charge cost}

\textcolor{black}{
\printnomenclature
}

\section{Introduction} \label{Introduction}

In the US and other parts of the world, the electricity tariff for most commercial consumers includes a monthly demand charge, which is calculated based on the highest average load requested from the grid, measured in kW, within a 15-minute interval of the monthly billing period~\cite{Luo2020}. Demand charge rates are usually one to two orders of magnitude higher than energy charge rates~\cite{hledik2014rediscovering,nazaripouya2019electric,ghosh2022effects}. Thus, demand charge management (DCM) based on economic cost minimization is attractive in \textcolor{black}{sizing,} day-ahead planning and real-time operation of microgrids. 

\textcolor{black}{The effective management of BESS or flexible loads can mitigate demand spikes for customers operating microgrids with high penetration of variable renewable energy (VRE) resources. Works such as~\cite{ogunmodede2021optimizing} describe the mathematical formulation and demonstrates the potential of  $\text {REopt Lite}^{\text {TM}}$~\cite{mishra2022computational,mishra2020nrel}, a web-based mixed integer programming optimization tool to optimally size and dispatch VRE, conventional generation and storage installations in a microgrid. However, as clarified by the authors in~\cite{ogunmodede2021optimizing}, models like REopt Lite work in a deterministic framework, optimizing over a representative year, where the weather (VRE resource data) and load for the entire year is known perfectly (or taken from a baseline) a-priori. While the deterministic framework makes REopt a powerful and computationally efficient tool for microgrid design, it is not adept for usage in real-time microgrid control where prediction horizons are much shorter than a year (typically $24-48$ h) with the possibility of forecasts being imperfect which needs to be accommodated by the system in real-time.}

While demand charges are calculated on a monthly basis, \textcolor{black}{for real-time microgrid control,} shifting the power requested from the grid (called demand) from higher demand hours to lower demand hours on a daily basis through a battery energy storage system (BESS) can lower monthly peak demand~\cite{hanna2014energy}. With the adoption of VRE sources, such as wind, solar, and electric vehicle (EV) charging stations, net loads have generally become more variable, and DCM with BESS has received much interest in the literature. BESS assisted DCM is applied in commercial and industrial buildings in~\cite{wang2017stochastic,raoufat2018model}, residential buildings in~\cite{luo2019optimal, Leadbetter2012, DeSalis2014}, utility-scale sites in~\cite{chen2021value, Kumar2018}, and at EV charging sites~\cite{ Chen2024, Yang2023, Lee2021, mohamed2020real} and are often combined with PV generation~\cite{wang2017stochastic,raoufat2018model,luo2019optimal}. 

One of the most common BESS control strategies for DCM is model predictive control (MPC), broadly subdivided into tracking and economic MPC. Tracking MPC computes control actions that minimize the deviation of the system operating point from a desired trajectory (or point) subject to input and state constraints over a finite prediction horizon. The inputs to the MPC are the latest state realizations (or observations) and forecasts of time-varying constraint parameters. At every time step, after computing the control actions for the entire prediction horizon, only the first control action is implemented. Then the system evolves to the next time step accordingly, and the optimization is repeated with updated inputs. However, the cost function and the desired trajectory in tracking MPC do not necessarily correspond to economic cost. Tracking MPC instead relies on the pre-computed trajectory to minimize demand charges. Defining a desired trajectory a-priori, for the system to follow in real-time for DCM, is challenging. 
 
Another way to optimize BESS dispatch for DCM is to directly use the economic cost function as the objective function over the prediction horizon, which is known as economic MPC (EMPC) \cite{Ellis2014}. This approach was used by the authors in~\cite{ghosh2023adaptive,mcclone2023hybrid} where the objective function is the electricity cost, composed of energy charges and demand charges. Note that, in reality, the demand charge is levied at the end of the month based on the maximum demand over the entire month, while the prediction horizon is shorter than a month (typically between $24$ \textcolor{black}{and} $48$~h). This mismatch of timescales can lead to greedy control actions toward the peak demands within the current prediction horizon, even when a much larger peak had already occurred during previous days of the month. These greedy control actions can result in BESS not having enough energy for the next high demand period, which can increase the peak demand, and the greedy control actions can also increase the BESS losses and degradation cost due to unnecessary BESS dispatch. Therefore, EMPC needs to keep track of the past peaks within the month to avoid greedy control actions. Previous works such as \cite{mcclone2023hybrid,ghosh2023adaptive} do not consider past peaks and implement the EMPC solutions naively in real-time, while other studies do not consider demand charges, which is unrealistic for many commercial consumers ~\cite{Parisio2016, Garifi2018}. 

Different approaches have been used in the literature to track previous peak demands. Some studies include a demand charge threshold (DCT) in the daily optimization algorithm~\cite{luo2019optimal, chen2021value, wang2017stochastic, Leadbetter2012, hanna2014energy, raoufat2018model, DeSalis2014, Yang2023, Kumar2018, Lee2021, Chen2024} and the DCT is adjusted in real time in~\cite{hanna2014energy, raoufat2018model, DeSalis2014, Yang2023, Kumar2018, Lee2021, Chen2024}. An initial value of DCT can be set at the beginning of the month, either based on historical data or expert knowledge. During the simulation throughout the month, DCT is included in the demand charge term of the cost function. After each time step, DCT is updated with the highest demand up to the current time step, storing the memory of the highest peak of the month. Then, only when the forecasted demand is higher than the DCT, the BESS is scheduled to dispatch to prevent higher peak demands.~\cite{hanna2014energy} first optimizes the persistence forecasted load to determine a daily DCT, which is then updated within the day in real time based on measurements.~\cite{raoufat2018model, DeSalis2014} estimate a monthly DCT from past measurements. While~\cite{raoufat2018model} adds a penalty for DCT violations to the cost function of an MPC,~\cite{DeSalis2014} updates the DCT during the day based on BESS constraints and demand forecast.~\cite{Chen2024} sets the initial monthly DCT to be zero and updates DCT in real time.~\cite{Yang2023} defines the demand charge cost term as the difference between the DCT computed up to the current time step and the predicted peak within the prediction horizon.~\cite{Kumar2018} incorporates a discount factor to split the total demand charge cost (based on a DCT) equally among the prediction horizon time points. Finally,~\cite{Lee2021} also applies the discount factor of~\cite{Kumar2018} in the demand charge cost term, but also considers an estimation of future peaks beyond the prediction horizon based on historical measurements. While the above traditional EMPCs with peak tracking reduce demand charge costs and battery cycling compared to traditional EMPCs without peak demand tracking, including a reference trajectory as an input to the EMPC formulation, can also lower electricity costs. A reference trajectory in EMPC has desirable closed-loop properties, such as asymptotic stability and performance guarantees~\cite{Risbeck2020}.

\textcolor{black}{The reference trajectory is a set of feasible states and control actions over the entire horizon. It can be derived from higher layers of long-term optimization for grid operators to design the objective functions for short-term optimizations. The long-term targets include economic, and safety objectives, and the short-term optimization considers the grid’s operational constraints while following the dispatch plan ~\cite{stai2020online, grammatikos2023design}. When the short-term cost function is known and the forecast for the PV generation and load is available, the reference trajectory consists of states and control actions from the first optimization layer} for the 
MPC to track and improve the system's economic performance~\cite{Angeli2015}.~\cite{Risbeck2020} first applied EMPC with a pre-defined monthly reference trajectory for DCM. A terminal cost and constraint are defined in the EMPC based on the reference trajectory. The EMPC also incorporated demand charge costs by including an auxiliary state. The auxiliary state tracks not only the previously achieved peaks but also tracks peaks within the prediction horizon. Under continuity and strict dissipativity assumptions~\cite{Risbeck2020}, the reference trajectory can guarantee stability and provide an upper bound to the average cost. These assumptions can be satisfied by the BESS. 

The reference trajectory benefits the EMPC in two aspects. (1) Knowledge about the system state and control actions from outside the prediction horizon is incorporated into the MPC. (2) The reference trajectory provides a first approach to the optimal control action that allows modifications on SOC regulation constraints when tracked by the EMPC afterward. A drawback of \cite{Risbeck2020} is the requirement of a full-month reference trajectory a-priori. A full-month reference trajectory is not practical because it requires knowledge about the MPC inputs (e.g., load, renewable energy generation) weeks in advance which is hard to obtain. In particular, renewable energy generation forecasts up to $24$ or $48$~h in advance are more practical.

\begin{figure} 
\centering
\includegraphics[clip,width=\columnwidth]{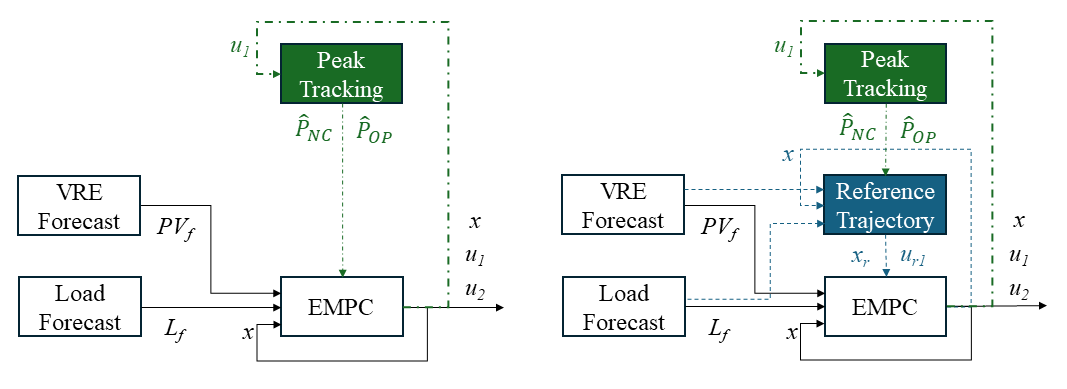}
\vspace{-1.5em}
\caption{EMPC frameworks that include load and VRE forecasts, peak tracking, and reference trajectory methods for traditional (left) and proposed (right) EMPC. \textcolor{black}{The following variables are presented: SOC of the BESS ($x$), charging/discharging power of the BESS ($u_2$), demand imported from the grid ($u_1$), forecasted load ($L_f$), forecasted PV generation ($PV_f$), NC and OP peak demands ($\hat{P}_{\text{NC}}$ and $\hat{P}_{\text{OP}}$), SOC of the BESS estimated as the reference trajectory ($x_r$), and demand imported from the grid estimated as the reference trajectory ($u_{r1}$).}}
\label{EMPCscheme}
\vspace{-1em}
\end{figure}

Inspired by~\cite{Risbeck2020}, we propose an optimal BESS scheduling model for DCM based on an EMPC algorithm as shown on the right side of Fig.~\ref{EMPCscheme}. In the proposed EMPC algorithm, the reference trajectory is updated online and then used in the terminal cost and constraint. The work involves two stages: ($i$) In the reference stage, we obtain a reference trajectory consisting of BESS state of charge (SOC) and control actions by solving an optimization problem, which includes a $50$\% SOC low threshold as a SOC regulation constraint, with and without peak demand tracking over a practical reference trajectory horizon (24 and 48~h); and ($ii$) In the MPC stage, we run an EMPC with a terminal cost and a terminal SOC constraint over a prediction horizon using the reference trajectory from ($i$) as input to the EMPC. Finally, we implement only the first control action obtained from ($ii$), allow the system to evolve, and repeat the process from ($i$). The performance of the proposed EMPC algorithm is compared with a traditional EMPC without a reference trajectory (EMPC framework presented on the left side of Fig.~\ref{EMPCscheme}). Both algorithms are compared under different configurations: ($i$) with and without incorporating the memory of the monthly peak demand, ($ii$) for $24$ and $48$~h horizon lengths, and ($iii$) with shrinking and rolling horizons.

The contributions of this work are as follows:
\begin{enumerate} 
\item Contrary to \cite{Risbeck2020} where the reference trajectory is assumed a-priori for the entire month, we propose a practical reference trajectory horizon equal to $24$ or $48$~h, in line with real-world applications. The reference trajectory is updated at each time step by solving an optimal control problem where peak demand tracking compensates for the timescale mismatch between the reference trajectory horizon and the demand charge billing cycle. 

\item We compare the performance of traditional versus proposed EMPC with and without peak demand tracking for shrinking and rolling horizons using real data. Both $24$ and $48$~h horizon lengths are considered. Simulation results demonstrate that extending the reference trajectory horizon improves the proposed EMPC performance. Furthermore, the proposed EMPC shows superior performance than the traditional EMPC implemented over a $48$~h reference and prediction horizons, with and without peak demand tracking and for shrinking and rolling horizons.

\item The proposed EMPC extends the work in \cite{Risbeck2020} to incorporate into the EMPC both non-coincident peak demand and on-peak demand charges, which resembles common electricity tariffs. \cite{Risbeck2020} only considered non-coincident demand charges.

\end{enumerate}

The rest of the paper is organized as follows. Section~\ref{Methodology} presents the traditional and proposed EMPC algorithms, as well as the optimal control problem solved to generate the reference trajectory in the proposed EMPC. Section~\ref{CaseStudy} presents the simulated case study. Simulation results are discussed in Section~\ref{results}. Finally, Section~\ref{conclusion} concludes the paper.

\section{Methodology}\label{Methodology}

\subsection{Time scheme}\label{TimeScheme} 

\begin{figure*} 
\centering
\includegraphics[trim={1cm 3cm 1cm 1.5cm},width=\textwidth]{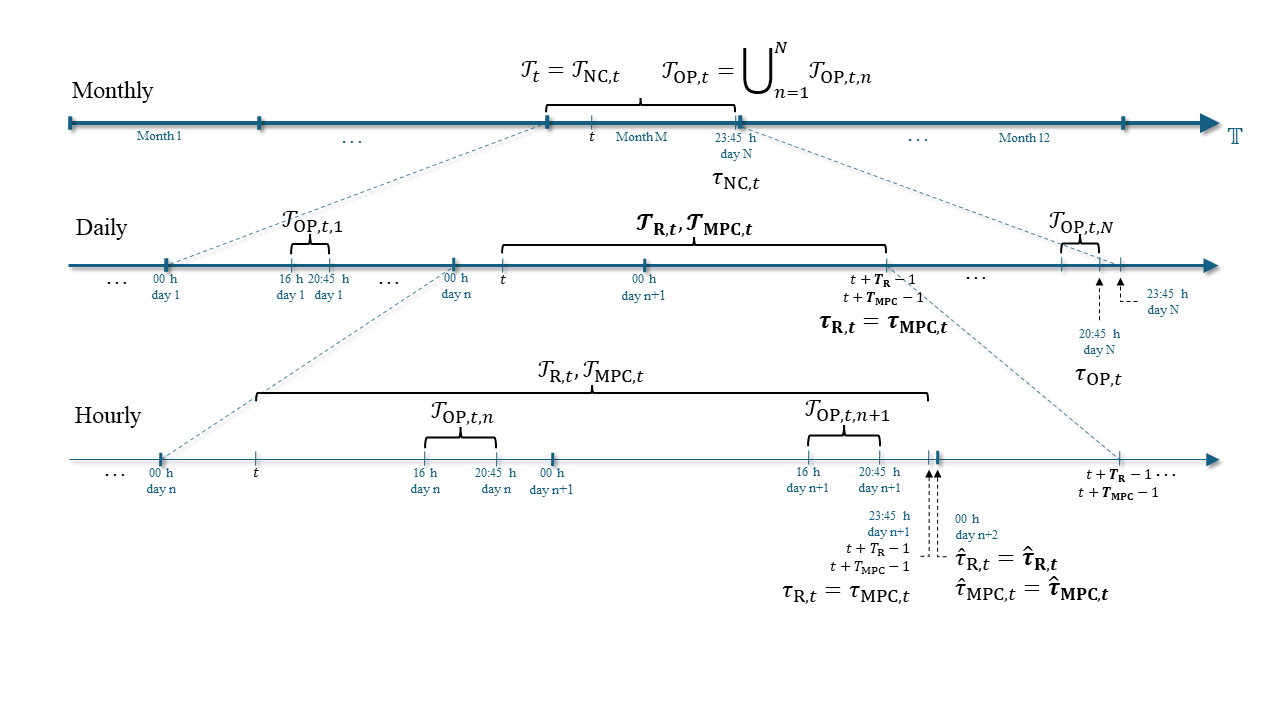}
\caption{Temporal schematic example for a $48$~h prediction horizon together with a $48$~h reference horizon, \textcolor{black}{both starting at time $t$, which falls on} day $n$ of a month $M$ with $N$ days. \textcolor{black}{The set of time points for the month $M$ is denoted by $\mathcal{T}_t$, with $\mathcal{T}_{\text{NC}, t}$ and $\mathcal{T}_{\text{OP}, t}$ defining the subsets corresponding to NC and OP demand charge periods, respectively. The final time points for these periods are represented as $\tau_{\text{NC},t}$ and $\tau_{\text{OP},t}$. The set $\mathcal{T}_{\text{OP}, t}$ is the union of the daily subset of OP demand charge time points $\mathcal{T}_{\text{OP}, t,n}$. Starting from $t$, the set of all time points in the reference horizon of length $T_{\text{R}}$ are defined in $\mathcal{T}_{\text{R},t}$, with its final time point called $\tau_{\text{R},t}$ and $\hat{\tau}_{\text{R},t}$ marking the time point at which the $50$\% SOC low threshold is placed. Similarly, the prediction horizon starting at the time point $t$, with length $T_{\text{MPC}}$, is represented by $\mathcal{T}_{\text{MPC},t}$. The final time point in this horizon is $\tau_{\text{MPC},t}$, and $\hat{\tau}_{\text{MPC},t}$ identifies the time point at which the $50$\% SOC low threshold is placed.} Bold and unbold time parameters $\mathcal{T}_{\text{MPC},t}$, $\mathcal{T}_{\text{R},t}$,$T_{\text{MPC}}$, $T_{\text{R}}$, $\tau_{\text{MPC},t}$, $\tau_{\text{R},t}$, $\hat{\tau}_{\text{MPC},t}$, and $\hat{\tau}_{\text{R},t}$ designate rolling and shrinking horizons, respectively.
}
\label{TimeDomainFigure}
\end{figure*}

Let $\mathbb{T}$ be the set of time points as shown in Fig.~\ref{TimeDomainFigure}, where $t\in \mathbb{T}$ is a time point. Since the demand charge horizon is equal to the entire month, for a given $t$, $\mathcal{T}_t$ is defined as the set of all time points in the month that $t$ belongs to. This work considers two demand charge peaks: (i) the monthly non-coincident (NC) peak demand, which is the highest average power demand in any 15-minute interval of the month, and (ii) the on-peak (OP) peak demand, which is the highest average power demand of a 15-minute interval during $16$:$00$~h to $21$:$00$~h of the month. $\mathcal{T}_{\text{NC}, t}$ and $\mathcal{T}_{\text{OP},t}$ are defined as subsets of $\mathcal{T}_t$ containing all time points corresponding to NC and OP demand charge periods for the entire month. \textcolor{black}{Note that $\mathcal{T}_{\text{OP},t}$ is a disconnected set, as shown in Fig.~\ref{TimeDomainFigure}}. The time points corresponding to the prediction horizon at the current MPC step $t$ are collected in $\mathcal{T}_{\text{MPC},t}$ and $T_{\text{MPC}}$ is the prediction horizon length. Similarly, the time points corresponding to the reference trajectory horizon (called reference horizon) are collected in $\mathcal{T}_{\text{R},t}$ and $T_{\text{R}}$ is the reference horizon length. The time points $\tau_{\text{NC},t}$ and $\tau_{\text{OP},t}$ correspond to the final time points of $\mathcal{T}_{\text{NC},t}$ and $\mathcal{T}_{\text{OP},t}$, respectively, and are used to reset the demand charges at the end of the month. Lastly, $\tau_{\text{MPC},t}$ and $\tau_{\text{R},t}$ are the final time points of $\mathcal{T}_{\text{MPC},t}$ and $\mathcal{T}_{\text{R},t}$, respectively.

\textcolor{black}{Note that the reference horizon $\mathcal{T}_{\text{R},t}$ and the prediction horizon $\mathcal{T}_{\text{MPC},t}$ represent different sets of time points. While both horizons may start at the same time point $t$, $\mathcal{T}_{\text{R},t}$ corresponds to the time period over which the reference trajectory is computed (further explanation provided in Section~\ref{EMPCsubsection}). Conversely, $\mathcal{T}_{\text{MPC},t}$ defines the prediction horizon over which an MPC algorithm is defined. As expected, $\mathcal{T}_{\text{R},t}$ and $\mathcal{T}_{\text{MPC},t}$ will correspond to the same time period for cases where both have the same length. Note that we refer to the MPC horizon as the prediction horizon.}

\subsection{Electricity cost}\label{ElectricityCost} 

A microgrid (MG) containing a BESS and a PV plant schedules the BESS to minimize the cost of electricity purchased from the grid. The electricity cost is assessed monthly and includes the energy cost and two peak demand charge costs. \textcolor{black}{At each $t \in \mathbb{T}$,} the energy cost is formulated as 
\begin{align} \label{energycost}
\mathcal{C}_1 \bigl(u_1(\textcolor{black}{t}),u_2(\textcolor{black}{t}) \bigl) = R_{\text{EC}} \Delta{T} \Bigl( u_1(\textcolor{black}{t}) + \frac{(1-\eta)}{2} |u_2(\textcolor{black}{t})| \Bigl), 
\end{align}
\textcolor{black}{where} ${R_{\text{EC}}}$ is the energy charge rate, $\Delta{T}$ is the time step size, and ${\eta}$ is the \textcolor{black}{roundtrip} efficiency of the BESS. The first term in $\eqref{energycost}$ is the cost of the energy purchased from the grid excluding BESS losses, with ${u_1}$ as the demand or power imported from the grid. The second term represents the BESS losses, with ${u_2}$ as the charging/discharging power of the BESS ($u_{2}>0$ when the BESS is discharging and $u_{2}<0$ when the BESS is charging). \textcolor{black}{The energy loss at each charging and discharging step is estimated as $R_{\text{EC}} \Delta{T} (1-\eta)|u_2(t)| / 2$. The factor of $1/2$ is included to correct for double-counting of energy losses over one charging and discharging cycle, as the implemented formulation does not differentiate between charging and discharging steps.} It is assumed that the utility allows the MG to sell back electricity at the energy charge rate.

The demand charge cost term is given by,
\begin{align} 
\mathcal{C}_2 &\bigl(\textcolor{black}{\bm{u_1}},\hat{P}_{\text{NC}}(t),\hat{P}_{\text{OP}}(t),t\textcolor{black}{,\tau_{e,t}} \bigl) = \label{DCcost}\\
&R_{\text{NC}} \max \{P_{\text{NC}}(\textcolor{black}{\bm{u_1}},t\textcolor{black}{,\tau_{e,t}}),\hat{P}_{\text{NC}}(t)\} + \nonumber\\ 
&R_{\text{OP}} \max \{P_{\text{OP}}(\textcolor{black}{\bm{u_1}},t\textcolor{black}{,\tau_{e,t}}),\hat{P}_{\text{OP}}(t)\},\nonumber 
\end{align}
where ${R_{\text{NC}}}$ and ${R_{\text{OP}}}$ are the NC and OP demand charge rates, respectively. \textcolor{black}{$\bm{u_1}$ is a vector where each element corresponds to $u_1(t)$ for all $t\in \mathbb{T}$, and $\tau_{e,t} \in \mathbb{T}$ is the final time point of a specific (future) horizon starting at $t$ where $\mathcal{C}_2$ is evaluated.} In~\eqref{DCcost}, $\hat{P}_{\text{NC}}$ and $\hat{P}_{\text{OP}}$ are the DCTs that track the NC and OP peak demands \textcolor{black}{($u_1$ peaks)} observed up to time $t$, respectively. Similarly, $P_{\text{NC}}$ and $P_{\text{OP}}$ are the NC and OP peak demands in \textcolor{black}{the horizon defined by $t$ and $\tau_{e,t}$}. These peaks are defined as,
\begin{subequations}
\begin{equation} \label{MoPeakTrackingNC} 
\begin{aligned} 
\hat{P}_{\text{NC}}(t+1) &= \textcolor{black}{\bigl(1-\sigma_{\text{NC}}(t+1) \bigl) \max \{ \hat{P}_{\text{NC}}(t), u_{1}(t) \}}  \\ 
\end{aligned}
\end{equation}
\begin{equation} \label{MoPeakTrackingOP}
\begin{aligned} 
\hat{P}_{\text{OP}}(t+1) &= \begin{cases} 
\textcolor{black}{\bigl(1-\sigma_{\text{OP}}(t+1) \bigl) \max \{ \hat{P}_{\text{OP}}(t), u_{1}(t)\} } \\ 
\quad \text{ if } t \in \textcolor{black}{\mathcal{T}_{\text{OP},t}} \\
\textcolor{black}{\bigl(1-\sigma_{\text{OP}}(t+1) \bigl) \max \{ \hat{P}_{\text{OP}}(t), 0\} } \\  
\quad \text{ otherwise},
\end{cases}
\end{aligned}
\end{equation}
\end{subequations}
\begin{subequations}
\begin{align}
P_{\text{NC}}(\textcolor{black}{\bm{u_1}},t\textcolor{black}{,\tau_{e,t}}) &= \textcolor{black}{\max_{k \in \{t,\dots,\tau_{e,t}\} \cap \mathcal{T}_{\text{NC},t}} u_1(k)} \label{PeakTrackingNC} \\
P_{\text{OP}}(\textcolor{black}{\bm{u_1}},t\textcolor{black}{,\tau_{e,t}}) &= \begin{cases} 
\max_{\textcolor{black}{k} \in \{t,\dots,\tau_{e,t}\} \cap \mathcal{T}_{\text{OP},t}} u_1(\textcolor{black}{k}) \\
\quad \text{ if } \{t,\dots,\tau_{e,t}\} \cap \mathcal{T}_{\text{OP},t} \neq \emptyset \\
0 \quad \text{otherwise}, \label{PeakTrackingOP} 
\end{cases}
\end{align}
\end{subequations}
where \textcolor{black}{$\{t,\dots,\tau_{e,t}\} \cap \mathcal{T}_{\text{NC},t}$ ($\{t,\dots,\tau_{e,t}\} \cap \mathcal{T}_{\text{OP},t}$) is the set of time points between $t$ and $\tau_{e,t}$ that also belong to the NC (OP) demand charge periods}. The demand charges consist of NC and OP charges, however,~\eqref{DCcost} can easily be modified to fit a different cost structure, e.g.~\cite{Yang2023, Kumar2018}. Assuming that neither expert knowledge nor historical demand data are available to determine an initial threshold at the beginning of the month for peak demands, $\hat{P}_{\text{NC}}(0)$ and $\hat{P}_{\text{OP}}(0)$ are set to zero. $\sigma_{\text{NC}}$ and $\sigma_{\text{OP}}$ are binary variables used to reset $\hat{P}_{\text{NC}}(t)$ and $\hat{P}_{\text{OP}}(t)$ at the beginning of each month
\begin{subequations}
\begin{equation} \label{sigmaequation NC}
\begin{aligned}
\sigma_{\rm NC}(t) &= \begin{cases}
1  & \text{if }  t \in \mathbb{T}_{\rm NC} \\
0 & \text{otherwise},
\end{cases}
\end{aligned}
\end{equation}
\begin{equation} \label{sigmaequation OP}
\begin{aligned}
\sigma_{\rm OP}(t) &= \begin{cases}
1  & \text{if }  t \in \mathbb{T}_{\rm OP} \\
0 & \text{otherwise},
\end{cases}
\end{aligned}
\end{equation}
\end{subequations}
where $\mathbb{T}_{\rm NC} = \lbrace0\rbrace \bigcup \lbrace\tau_{\text{NC},t}+1 | t \in \mathbb{T} \rbrace$ ($\mathbb{T}_{\rm OP} = \lbrace0\rbrace \bigcup \lbrace\tau_{\text{OP},t}+1 | t \in \mathbb{T} \rbrace$) is the set of time points that correspond to the beginning of a month of the NC (OP) demand charge horizon.

\subsection{Traditional EMPC without peak demand tracking}\label{TraditionalNT}

\textcolor{black}{Let the state $x$ represent the SOC of the BESS. At $t \in \mathbb{T}$ with the state $x(t)$,} the traditional EMPC formulation without peak demand tracking is given by,
\begin{subequations} \label{TradNT}
\begin{align} 
\min_{ u_1,u_2} \quad & \sum_{k=t}^{\tau_{\text{MPC},t}} \mathcal{C}_{1} \bigl(u_1(k),u_2(k) \bigl) + \nonumber \\
&\mathcal{C}_{2} \bigl( \textcolor{black}{\bm{u_1}},0,0,t\textcolor{black}{,\tau_{\text{MPC},t}} \bigl), \label{TradNTObjFn}\\  
\textrm{s.t.} \quad x(k+1) &= x(k) - \frac{u_2(k)}{\text{BESS}_{en}} \Delta{T}, \label{SOCtime}\\
L_{f}(k) &= \text{PV}_{f}(k) + u_{2} (k) + u_{1}(k), \label{power balance}\\  
-\overline{u}_{2} &\leq u_{2} (k)\leq \overline{u}_{2}, \label{BESSpower} \\
\underline{x} &\leq x(k)\leq \overline{x}, \label{BESSSOC}\\
& \textcolor{black}{\forall\,{k \in \mathcal{T}_{\text{MPC},t}}}, \nonumber\\ 
x(\hat{\tau}_{\text{MPC},t}) &\geq 0.5. \label{FinalSOC}
\end{align} 
\end{subequations}
\textcolor{black}{where $k$ is used to denote the prediction horizon $\mathcal{T}_{\text{MPC},t}=\{t,...,\tau_{\text{MPC},t}\}$.} Fig.~\ref{TimeDomainFigure} shows the corresponding prediction horizon temporal scheme. The objective function \eqref{TradNTObjFn} minimizes the total energy cost and the demand charges assessed over the prediction horizon of length $T_{\text{MPC}}$. The constraint~\eqref{SOCtime} is the state update equation for the SOC $x$, where $\text{BESS}_{\rm en}$ is the BESS energy capacity. Constraint \eqref{power balance} is the load balance equation, which ensures that the MG satisfies the forecasted load, $L_{f}$. $\text{PV}_{f}$ is the forecasted PV generation. Constraints \eqref{BESSpower} and \eqref{BESSSOC} represent BESS power limits and SOC limits, where $\overline{u}_{2}$ is the BESS power capacity and $\overline{x}$ and $\underline{x}$ are the maximum and minimum SOC limits. Constraint \eqref{FinalSOC} defines a $50$\% SOC low threshold to be satisfied at time $\hat{\tau}_{\text{MPC},t}$.

\textcolor{black}{Note that BESS losses when $\eta<1$ are considered only in~\eqref{energycost}. The load balance in~\eqref{power balance} and the state update in~\eqref{SOCtime} assume the BESS is $100$\% efficient, as in~\cite{Kumar2018, ghosh2023adaptive}, to maintain convexity. The work presented in~\cite{ghosh2023adaptive} implements a traditional EMPC without peak demand tracking.}

\textcolor{black}{The MG model presented above has minor shortcomings that are important to address for future implementations. First, the model does not restrict the power imported from the grid, which can be addressed as an upper (and lower) limit constraint on $u_1$. Second, assigning different energy rates for power imported from and exported to the grid represents another improvement, which can be achieved by incorporating separate energy cost terms for positive and negative values of $u_1$ in~\eqref{energycost}. Finally, the model does not account for fossil fuel generation units, which could be included by adding their quadratic cost to the objective function.}

\subsection{Traditional EMPC with peak demand tracking}\label{TraditionalWT}

The traditional EMPC presented above can also include memory of the previously achieved peaks by incorporating $\hat{P}_{\text{NC}}(t)$ and $\hat{P}_{\text{OP}}(t)$ from~\eqref{MoPeakTrackingNC} and~\eqref{MoPeakTrackingOP} into $\mathcal{C}_2$. \textcolor{black}{At $t \in \mathbb{T}$ with the state $x(t)$,} the resulting traditional EMPC algorithm with peak demand tracking is defined as,
\begin{equation} 
\label{TradWT}
\begin{aligned} 
\min_{u_1,u_2} & \sum_{k=t}^{\tau_{\text{MPC},t}} \mathcal{C}_{1} \bigl(u_1(k),u_2(k) \bigl) + \\
&\mathcal{C}_{2} \bigl(\textcolor{black}{\bm{u_1}},\hat{P}_{\text{NC}}(t),\hat{P}_{\text{OP}}(t),t\textcolor{black}{,\tau_{\text{MPC},t}} \bigl), \\
\textrm{s.t.} \quad & ~\eqref{SOCtime},~\eqref{power balance},~\eqref{BESSpower},~\eqref{BESSSOC}, \\
& \textcolor{black}{\forall\,{k \in \mathcal{T}_{\text{MPC},t}}}, \\   
&~\eqref{FinalSOC}.
\end{aligned} 
\end{equation}

\subsection{SOC low threshold} \label{SOClowthreshold}

The timescale mismatch between the monthly demand charge billing cycle and the real-world prediction horizon length of $24$ to $48$~h requires the addition of a SOC regulation constraint, which is implemented as a $50$\% SOC low threshold. This constraint ensures that the BESS has sufficient energy at the beginning of each day as the actual monthly peak demand could occur on any day of the month. The SOC low threshold at the beginning of each day would not be needed if the prediction horizon spanned the entire month. In that case, the SOC would be allowed to reach any value at the end of each day to minimize monthly peak demands.

However, enforcing the SOC low threshold limits the BESS flexibility within the prediction horizon. For example, when the prediction horizon is $24$~h, the SOC low threshold will be placed at most $24$~h ahead. Lesser BESS flexibility will be available for peak demand reduction within the prediction horizon when compared to cases without a low threshold. Therefore, where the end of the prediction horizon is defined, and where the $50$\% SOC low threshold is placed, play a crucial role in the BESS flexibility to reduce demand charges. The $50$\% SOC low threshold could also ensure that the BESS has sufficient energy to balance forecast errors. Forecast errors are not considered in this work and will be explored in a future publication.

We consider two approaches for the prediction horizon length and the placement of the $50$\% SOC low threshold: shrinking and rolling prediction horizon. Shrinking the prediction horizon implies that the final time point of the prediction horizon, $\tau_{\text{MPC},t}$, is fixed at the end of a day and it does not advance forward to the next day. This means that $\tau_{\text{MPC},t}$ is defined at the beginning of the day and remains fixed until $t$ advances to the next day. Also, the $50$\% SOC low threshold is placed at the end of the prediction horizon, i.e., $\hat{\tau}_{\text{MPC},t}=\tau_{\text{MPC},t}+1$.

On the other hand, the horizon length is kept constant for rolling prediction horizons such that the last horizon time point advances at every MPC step, and thus $\tau_{\text{MPC},t}$ changes at every MPC step. Under the rolling prediction horizon, the $50$\% SOC low threshold is placed at the last midnight time point included in the prediction horizon and moves to midnight of the next day when $t$ advances to the next day. \textcolor{black}{Under this setting, the $50$\% SOC low threshold will not prevent the BESS from being fully discharged at time points after the constraint, but it will still limit BESS operation during earlier time points within the prediction horizon.}

\subsection{Proposed EMPC}\label{EMPCsubsection}

\subsubsection{MPC Stage}\label{EMPCformulation} 

In the MPC stage of the proposed EMPC \textcolor{black}{algorithm, the mathematical formulation developed by~\cite{Risbeck2020} is adapted to incorporate both demand charges (NC and OP) and the online reference trajectory. First,} the system model is augmented with two auxiliary state variables, $y_{\rm NC}$ and $y_{\rm OP}$, to track the two demand charge peaks \textcolor{black}{within the prediction horizon $\mathcal{T}_{\text{MPC},t}$ starting at time $t$ and denoted using $k$. \textcolor{black}{The temporal scheme of the prediction horizon is shown in Fig.~\ref{TimeDomainFigure}.} The variables $y_{\rm NC}$ and $y_{\rm OP}$} are defined as,
\begin{subequations} \label{Ally_update}
\begin{align} 
y_{\rm NC}(k+1) &= \max\{ \bigl(1-\sigma_{\rm NC}(k) \bigl)y_{\rm NC}(k),u_{1}(k) \}\label{yNC_update}, \\
y_{\rm OP}(k+1) &= \begin{cases} 
\max\{ \bigl(1-\sigma_{\rm OP}(k) \bigl)y_{\rm OP}(k),u_{1}(k) \} \\
  \quad \text{ if } k \in \mathcal{T}_{\text{OP},t} \\
\max\{ \bigl(1-\sigma_{\rm OP}(k) \bigl)y_{\rm OP}(k),0 \} \\
  \quad \text{ otherwise},
\end{cases} \label{yOP_update}
\end{align}
\end{subequations}
for $k>t$ \textcolor{black}{, and where $\sigma_{\text{NC}}$ and $\sigma_{\text{OP}}$ are used to reset $y_{\rm NC}$ and $y_{\rm OP}$ at the beginning of each month. $y_{\rm OP}$ is defined as a piecewise function because it only considers demand values $u_1$ registered during the OP demand charge set $\mathcal{T}_{\text{OP},t}$, which is disconnected}. Both auxiliary variables are initialized at the start of the prediction horizon ($k=t$) with the previously observed peaks ($\hat{P}_{\text{NC}}(t)$ and $\hat{P}_{\text{OP}}(t)$) up to $t$. Specifically,
\begin{align}
y_{\rm NC}(t) = \hat{P}_{\text{NC}}(t), \quad
y_{\rm OP}(t) = \hat{P}_{\text{OP}}(t) \label{yNC_OP_initial}.
\end{align}

The cost function consists of a stage cost and a terminal cost. The stage cost accounts for the energy cost and the demand charges incurred at the end of the month. Denoting $z(k)= \bigl(x(k),y_{\rm NC}(k),y_{\rm OP}(k) \bigl)^\top$, the stage cost is defined as,  
\begin{multline} \label{stagecost}
L \bigl(z(k),u_{1}(k),u_{2}(k),k \bigl) = \mathcal{C}_1 \bigl(u_{1}(k),u_{2}(k) \bigl) + \\ R_{\text{NC}} \sigma_{\rm NC}(k) y_{\rm NC}(k) + R_{\text{OP}} \sigma_{\rm OP}(k) y_{\rm OP}(k),
\end{multline}
where the second and third terms correspond to the demand charge costs and are added to the stage cost only when the end of the month is within the prediction horizon, that is, when any time point in $\mathbb{T}_{\text{NC}}$ or $\mathbb{T}_{\text{OP}}$ belongs to $\mathcal{T}_{\text{MPC},t}$.

The terminal cost uses the reference trajectory to account for demand charges of time points that correspond to a month whose final time point is not within the prediction horizon. Let $z_r(\textcolor{black}{k'}) = \bigl( x_r(\textcolor{black}{k'}),u_{r1}(\textcolor{black}{k'}),u_{r2}(\textcolor{black}{k'}) \bigl)$ be the reference trajectory over the reference horizon $\mathcal{T}_{\text{R},t}$\textcolor{black}{, which is denoted using $k'$ and starts at time $k'=t$. Then, for each $k \in \mathcal{T}_{\text{MPC},t} \cup \{\tau_{\text{MPC},t}+1\}$,} we define \textcolor{black}{the reference trajectory peaks as follows,}
\begin{subequations} \label{Allyparameters}
\begin{align}
\hat{y}_{\rm NC}(\textcolor{black}{k,\hat{k}}) &= \max_{\textcolor{black}{k'} \in \mathcal{T}_{\text{NC},\textcolor{black}{k}-1} \cap \mathcal{T}_{\text{R},\textcolor{black}{\hat{k}}}} \{u_{r1}(\textcolor{black}{k'}),\hat{P}_{\text{NC}}(\textcolor{black}{\hat{k}})\}, \label{yparameter1NC}\\
\check{y}_{\rm NC}(\textcolor{black}{k},\textcolor{black}{\hat{k}}) &= \bigl(1-\sigma_{\text{NC}}(\textcolor{black}{k}) \bigl) \max_{\textcolor{black}{k'} \in \mathcal{T}_{\text{NC},\textcolor{black}{k}} \cap \mathcal{T}_{\text{R},\textcolor{black}{\hat{k}}}\textcolor{black}{,k' \geq k}} u_{r1}(\textcolor{black}{k'}), \label{yparameter2NC}\\
\hat{y}_{\rm OP}(\textcolor{black}{k},\textcolor{black}{\hat{k}}) &= \max_{\textcolor{black}{k'} \in \mathcal{T}_{\text{OP},\textcolor{black}{k}-1} \cap \mathcal{T}_{\text{R},\textcolor{black}{\hat{k}}}} \{u_{r1}(\textcolor{black}{k'}),\hat{P}_{\text{OP}}(\textcolor{black}{\hat{k}})\}, \label{yparameter1OP}\\
\check{y}_{\rm OP}(\textcolor{black}{k},\textcolor{black}{\hat{k}}) &= \bigl(1-\sigma_{\rm OP}(\textcolor{black}{k}) \bigl) \max_{\textcolor{black}{k'} \in \mathcal{T}_{\text{OP},\textcolor{black}{k}} \cap \mathcal{T}_{\text{R},\textcolor{black}{\hat{k}}}\textcolor{black}{,k' \geq k}} u_{r1}(\textcolor{black}{k'}), \label{yparameter2OP} 
\end{align} 
\end{subequations}
where \textcolor{black}{$\hat{y}_{\rm NC}(k,\hat{k})$ ($\hat{y}_{\rm OP}(k,\hat{k})$) is the maximum of the peak power imported from the grid estimated by the reference trajectory, the $u_{r1}$ peak, for NC (OP) time points and the $u_1$ peak until time $\hat{k}$ for NC (OP) time points, which is stored in $\hat{P}_{\text{NC}}(\hat{k})$ ($\hat{P}_{\text{OP}}(\hat{k})$). $\hat{k}$ is the time point where the reference trajectory was computed, with $\hat{k} < k$.} \textcolor{black}{Furthermore, the time shift in~\ref{yparameter1NC} (\ref{yparameter1OP}) allows $\hat{y}_{\rm NC}(k,\hat{k})$ ($\hat{y}_{\rm OP}(k,\hat{k})$) to compute the peak for the demand charge period just ended, that is, when $k \in \mathbb{T}_{\rm NC}$ ($k \in \mathbb{T}_{\rm OP}$) as explained in~\cite{Risbeck2020}.} $\check{y}_{\rm NC}(\textcolor{black}{k},\textcolor{black}{\hat{k}})$ ($\check{y}_{\rm OP}(\textcolor{black}{k},\textcolor{black}{\hat{k}})$) corresponds to the $u_{r1}$ peak for the NC (OP) time points within $\mathcal{T}_{\text{R},\textcolor{black}{\hat{k}}}$ \textcolor{black}{but only after the time point $k$.  Given that the set $\mathcal{T}_{\text{OP},k}$ is disconnected, whenever $\mathcal{T}_{\text{OP},k} \cap \mathcal{T}_{\text{R},\hat{k}}$ is an empty set in shrinking horizon cases, a value of zero is assigned as the maximum of $u_{r1}$.}

In contrast with~\cite{Risbeck2020}, where the reference horizon starts at the beginning of the month and the reference trajectory is generated in one shot for the entire month, in this work \textcolor{black}{the reference trajectory} $\mathcal{T}_{\text{R},\textcolor{black}{\hat{k}}}$ \textcolor{black}{is updated at every MPC step, beginning at time $\hat{k}$ and generated for a shorter horizon. Furthermore, the original reference trajectory peak formulations from~\cite{Risbeck2020} were accordingly adapted in~\eqref{Allyparameters} to incorporate the online reference trajectory proposed.} 

The terminal cost is then defined as,
\begin{align}
V_f &\bigl(z(k),\textcolor{black}{k},\textcolor{black}{\hat{k}} \bigl) = \label{terminalcost} \\
&R_{\text{NC}} \bigl(\max \{y_{\rm NC}(k),\check{y}_{\rm NC}(\textcolor{black}{k},\textcolor{black}{\hat{k}})\} - \hat{y}_{\rm NC}(\textcolor{black}{k,\hat{k}}) \bigl) + \nonumber \\
&R_{\text{OP}} \bigl(\max \{y_{\rm OP}(k),\check{y}_{\rm OP}(\textcolor{black}{k},\textcolor{black}{\hat{k}})\} - \hat{y}_{\rm OP}(\textcolor{black}{k,\hat{k}}) \bigl).  \nonumber
\end{align}
With \eqref{terminalcost} as the terminal cost, there is no \textcolor{black}{incentive to reduce any of the peak demands below the peak predicted by the reference trajectory for time points beyond $k$}. This is because at \textcolor{black}{each time point $k$} a minimum value for the NC (OP) peak is defined by $\check{y}_{\rm NC}(\textcolor{black}{k,\hat{k}})$ ($\check{y}_{\rm OP}(\textcolor{black}{k,\hat{k}})$). In addition, when the end of the demand charge horizon is not included within the prediction horizon \textcolor{black}{starting at $\hat{k}$}, \eqref{terminalcost} penalizes increments in \textcolor{black}{the peak NC (OP) demand charge} $y_{\text{NC}}(k)$ ($y_{\text{OP}}(k)$) for $y_{\text{NC}}(k) > \check{y}_{\rm NC}(\textcolor{black}{k,\hat{k}})$ ($y_{\text{OP}}(k) > \check{y}_{\rm OP}(\textcolor{black}{k,\hat{k}})$) only by the excess amount with respect to $\hat{y}_{\rm NC}(\textcolor{black}{k,\hat{k}})$ ($\hat{y}_{\rm OP}(\textcolor{black}{k,\hat{k}})$).

The formulation developed by~\cite{Risbeck2020} also proposes a terminal constraint for the state $x$ based on the reference trajectory as,
\begin{equation}
\begin{aligned}
x(\tau_{\text{MPC},t}+1) &=x_r(\tau_{\text{MPC},t}+1). \label{terminalconstraint2}
\end{aligned}
\end{equation}
Finally, the MPC stage of the proposed EMPC is defined \textcolor{black}{at $t \in \mathbb{T}$ with the state $x(t)$} by,
\begin{equation} \label{EMPC}
\begin{aligned}
 \min_{u_1,u_2} \quad & \sum_{k=t}^{\tau_{\text{MPC},t}} L \bigl(z(k),u_{1}(k),u_{2}(k),k \bigl) + \\ &V_f \bigl(z(\tau_{\text{MPC},t}+1),\textcolor{black}{\tau_{\text{MPC},t}+1},\textcolor{black}{t} \bigl), \\
\textrm{s.t.} \quad &~\eqref{SOCtime},~\eqref{power balance},~\eqref{BESSpower},~\eqref{BESSSOC},~\eqref{yNC_update},~\eqref{yOP_update},\\
& \quad \textcolor{black}{\forall\,{k \in \mathcal{T}_{\text{MPC},t}}},\\ 
&~\eqref{yNC_OP_initial},~\eqref{terminalconstraint2}.\\
\end{aligned} 
\end{equation}

\noindent \textcolor{black}{Beside $t$, the terminal cost function $V_f$ in~ \eqref{EMPC} is evaluated using the states variables at time $\tau_{\text{MPC},t}+1$ and at that time point itself. This is done to incorporate demand charge costs into the formulation when the prediction horizon $\mathcal{T}_{\text{MPC},t}$ ends before the demand charge horizon, as the stage cost $L$ does not incorporate demand charges in such cases. In addition, it is important to note that all variables needed to compute $V_f\bigl(z(\tau_{\text{MPC},t}+1),\tau_{\text{MPC},t}+1,t \bigl)$ can be determined either from the reference trajectory $z_r(k') = \bigl( x_r(k'),u_{r1}(k'),u_{r2}(k') \bigl)$ (computed in advance) or from data collected within the prediction horizon $\mathcal{T}_{\text{MPC},t}$. The only exception occurs when the reference and prediction horizons are of equal length, in which case $\check{y}_{\rm NC}$ and $\check{y}_{\rm OP}$ cannot be computed. In such instances, the values computed by~\eqref{Allyparameters} at the previous time point $\tau_{\text{R},t}$ are reused for the subsequent time point $\tau_{\text{R},t}+1$.}

While the optimization in~\eqref{EMPC} will be also solved in shrinking and rolling horizon fashions, the traditional EMPC and the proposed EMPC differ in how they handle the SOC at the end of the prediction horizon $\mathcal{T}_{\text{MPC},t}$. The traditional EMPC algorithm implements a 50\% low SOC threshold at the time step $\hat{\tau}_{\text{MPC},t}$ given by \eqref{FinalSOC}, which corresponds to the terminal SOC constraint for shrinking horizons. The MPC stage of the proposed EMPC in \eqref{EMPC} does not include the $50$\% SOC low threshold as a constraint but the low threshold is instead enforced through the reference trajectory for both shrinking and rolling prediction horizons. The reference trajectory is an input to the MPC stage, and it is obtained by solving an optimization problem that includes the $50$\% SOC low threshold constraint as discussed next.

\subsubsection{Reference Trajectory without Peak Demand Tracking (Reference Stage)}
The reference trajectory $z_r(\textcolor{black}{k'})$ is computed at the reference stage of the proposed EMPC. Its temporal alignment is shown in Fig.~\ref{TimeDomainFigure}. The reference trajectory is an input to the MPC stage and is used as a basis for measuring the economic performance of the system. The reference trajectory without peak demand tracking is the sequence $\{(x_r(\textcolor{black}{k'}),u_{r1}(\textcolor{black}{k'}),u_{r2}(\textcolor{black}{k'}))\}_{\textcolor{black}{k' \in \mathcal{T}_{\text{R},t}}}$ that solves the following optimization problem over the reference horizon $\mathcal{T}_{\text{R},t}$,
\begin{subequations} \label{RefTrNT}
\begin{align}
\min_{u_{r1},u_{r2}} \quad & \sum_{\textcolor{black}{k'}=t}^{\tau_{\text{R},t}} \mathcal{C}_1 \bigl(u_{r1}(\textcolor{black}{k'}),u_{r2}(\textcolor{black}{k'}) \bigl) + \nonumber \\
&\mathcal{C}_{2} \bigl(\textcolor{black}{\bm{u_{r1}}},0,0,t\textcolor{black}{,\tau_{\text{R},t}} \bigl), \label{RefObjective}\\
\textrm{s.t.} \quad x_{r} (\textcolor{black}{k'}+1)&=x_{r}(\textcolor{black}{k'})-\frac{u_{r1}(\textcolor{black}{k'})}{\text{BESS}_{en}}\Delta{T}, \label{RefSOCtime} \\
L_{f}(\textcolor{black}{k'}) &= \text{PV}_{f}(\textcolor{black}{k'}) + u_{r2} (\textcolor{black}{k'}) + u_{r1} (\textcolor{black}{k'}), \label{Refpower balance} \\
\underline{u}_{2} &\leq u_{r2} (\textcolor{black}{k'})\leq \overline{u}_{2}, \label{RefBESSpower}\\
\underline{x} &\leq x_{r} (\textcolor{black}{k'})\leq \overline{x}, \label{RefBESSSOC}\\
& \textcolor{black}{\forall{k' \in \mathcal{T}_{\text{R},t}}}, \nonumber\\
x_{r} (t)&= \textcolor{black}{x(t)}, \label{RefInitialSOC_2}\\
x_{r} (\hat{\tau}_{\text{R},t})&\geq 0.5, \label{RefFinalSOC}
\end{align}
\end{subequations} 
where $\hat{\tau}_{\text{R},t}$ is the time point where the $50$\% SOC low threshold is placed \textcolor{black}{, and $\bm{u_{r1}}$ is a vector where each element corresponds to $u_{r1}(t)$ for all $t\in \mathbb{T}$}. Similar to the prediction horizons described above, the reference horizon can be shrinking or rolling. The constraint \eqref{RefFinalSOC} is placed at the end of the shrinking reference horizon ($\hat{\tau}_{\text{R},t}=\tau_{\text{R},t}+1$), while for the rolling reference horizon, the constraint \eqref{RefFinalSOC} is assigned to the last midnight time point within the reference horizon ($\hat{\tau}_{\text{R},t} \leq \tau_{\text{R},t}$).

\subsubsection{Reference Trajectory with Peak Demand Tracking (Reference Stage)}
A reference trajectory that incorporates previous peak demands is the control sequence $\{(x_r(\textcolor{black}{k'}),u_{r1}(\textcolor{black}{k'}),u_{r2}(\textcolor{black}{k'}))\}_{k' \in \mathcal{T}_{\text{R},t}}$ that solves the following optimization problem,
\begin{equation} \label{RefTrWT}
\begin{aligned}
\min_{u_{r1},u_{r2}} \quad & \sum_{\textcolor{black}{k'}=t}^{\tau_{\text{R},t}} \mathcal{C}_1 \bigl(u_{r1}(\textcolor{black}{k'}),u_{r2}(\textcolor{black}{k'}) \bigl) + \\
&\mathcal{C}_{2} \bigl(\textcolor{black}{\bm{u_{r1}}},\hat{P}_{\text{NC}}(t),\hat{P}_{\text{OP}}(t),t\textcolor{black}{,\tau_{\text{R},t}} \bigl), \\ 
\textrm{s.t.} \quad &~\eqref{RefSOCtime},~\eqref{Refpower balance},~\eqref{RefBESSpower},~\eqref{RefBESSSOC}, \\
& \textcolor{black}{\forall{k' \in \mathcal{T}_{\text{R},t}}}, \\
&~\eqref{RefInitialSOC_2},~\eqref{RefFinalSOC}, 
\end{aligned} 
\end{equation}
where compared to~\eqref{RefTrNT}, $\hat{P}_{\text{NC}}(t)$ and $\hat{P}_{\text{OP}}(t)$ have been included in $\mathcal{C}_2$ when computing demand charges in the objective function. Given the much shorter reference horizon ($24$-$48$~h) than the demand charge horizon (one month), a new reference trajectory is generated at every MPC step. 

\textcolor{black}{As was explained in Section~\ref{TimeScheme}, we refer to MPC horizon as the prediction horizon. Additionally, note that the reference trajectory is a set of feasible states and control actions, and the reference horizon $\mathcal{T}_{\text{R},t}$ and the prediction horizon $\mathcal{T}_{\text{MPC},t}$ define different sets of time points.}

\textcolor{black}{The implementation of the proposed EMPC considering a reference trajectory with peak demand tracking is summarized in Algorithm~\ref{alg:alg1}. The proposed EMPC considering a reference trajectory without peak demand tracking can also be summarized in Algorithm~\ref{alg:alg1}, but problem~\eqref{RefTrNT} will be solved in line $11$ instead of problem~\eqref{RefTrWT}.}

\begin{algorithm}
\caption{\textcolor{black}{Proposed EMPC with Peak Tracking}} \label{alg:alg1}
\begin{algorithmic}[1]
\STATE \textcolor{black}{Initialize $x(0)$.}
\textcolor{black}{
\FOR{$t \in \mathbb{T}$}
\IF{$t \in \mathbb{T}_{\rm NC}$}
\STATE Define the time set for the month $\mathcal{T}_{t}$;
\STATE Define the subsets corresponding to NC and OP demand charge periods $\mathcal{T}_{\text{NC}, t}$ and $\mathcal{T}_{\text{OP},t}$; 
\STATE Assign the final time points $\tau_{\text{NC},t}$ and $\tau_{\text{OP},t}$ of the subsets $\mathcal{T}_{\text{NC}, t}$ and $\mathcal{T}_{\text{OP},t}$.
\ENDIF
\item[] \textbf{Reference stage}
\STATE Observe system current state $x(t)$, and monthly peak tracking variables $\hat{P}_{\text{NC}}(t)$ and $\hat{P}_{\text{OP}}(t)$;
\STATE Define the reference horizon time set $\mathcal{T}_{\text{R}, t}$, and assign its final time point $\tau_{\text{R}, t}$ and the time point where the SOC low threshold is placed $\hat{\tau}_{\text{R}, t}$;
\STATE Forecast load $\{L_{f}(k')\}_{k' \in \mathcal{T}_{\text{R}, t}}$ and PV generation $\{PV_{f}(k')\}_{k' \in \mathcal{T}_{\text{R}, t}}$ over the reference horizon;
\STATE Solve the problem~\eqref{RefTrWT} to obtain the reference trajectory $\{x_r(k'),u_{r1}(k'),u_{r2}(k')\}_{k' \in \mathcal{T}_{\text{R}, t}}$;
\item[] \textbf{MPC stage}
\STATE Define prediction horizon time set $\mathcal{T}_{\text{MPC}, t}$, and assign its final time point $\tau_{\text{MPC}, t}$ and the time point where the SOC low threshold is placed $\hat{\tau}_{\text{MPC}, t}$;
\STATE Forecast load $\{L_{f}(k)\}_{k \in \mathcal{T}_{\text{MPC}, t}}$ and PV generation $\{PV_{f}(k)\}_{k \in \mathcal{T}_{\text{MPC}, t}}$ over the prediction horizon;
\STATE Compute the peaks of the reference trajectory $\hat{y}_{\rm NC}(k,\hat{k})$, $\check{y}_{\rm NC}(k,\hat{k})$, $\hat{y}_{\rm OP}(k,\hat{k})$ and $\check{y}_{\rm OP}(k,\hat{k})$ according to~\eqref{Allyparameters} with $\hat{k}=t$ and $t=\tau_{\text{MPC},t}+1$;
\STATE Observe system current state $x(t)$, peaks of the reference trajectory ($\hat{y}_{\rm NC}(\tau_{\text{MPC},t}+1,t)$, $\check{y}_{\rm NC}(\tau_{\text{MPC},t}+1,t)$, $\hat{y}_{\rm OP}(\tau_{\text{MPC},t}+1,t)$ and $\check{y}_{\rm OP}(\tau_{\text{MPC},t}+1,t)$), and monthly peak tracking variables $\hat{P}_{\text{NC}}(t)$ and $\hat{P}_{\text{OP}}(t)$;
\STATE Solve the problem~\eqref{EMPC} to obtain the optimal control $\{x(k),u_{1}(k),u_{2}(k)\}_{k \in \mathcal{T}_{\text{MPC}, t}}$ over the prediction horizon;
\STATE Save $x(t+1)$, $u_{1}(t)$, and $u_{2}(t)$;
\STATE Update monthly peak tracking variables $\hat{P}_{\text{NC}}(t+1)$ and $\hat{P}_{\text{OP}}(t+1)$ according to~\eqref{MoPeakTrackingNC} and~\eqref{MoPeakTrackingOP}. 
\ENDFOR
}
\end{algorithmic}
\end{algorithm}

\section{Case Study} \label{CaseStudy}

\subsection{Microgrid at the Port of San Diego} \label{MicrogridPSD}

The MG at the Port of San Diego (latitude $32.69^{\circ}$, longitude $-117.14^{\circ}$) consists of a $700$~kW PV plant and a BESS system. The BESS has an energy capacity $\text{BESS}_{\rm en}=2,500$~kWh and power capacity $\overline{u}_{2}=700$~kW. The BESS charging/discharging efficiency (${\eta}$) is $80$\%, and the maximum ($\overline{x}$) and minimum ($\underline{x}$) SOC limits the BESS is allowed to reach are $0.8$ and $0.2$, respectively. The electricity cost of the MG considers a constant energy rate ($R_{\text{EC}}$) of \$$0.1$/kWh, and NC and OP demand charge rates ($R_{\text{NC}}$ and $R_{\text{OP}}$) of \$$24.48$/kWh and \$$19.19$/kWh, respectively. \textcolor{black}{These are non-residential rates from 2020 (SDG\&E AL-TOU tariff)~\cite{SDGE1}. $R_{\text{OP}}$ ($R_{\text{EC}}$) is the average between the winter and summer OP demand charge (Off-Peak energy) rates}. While energy rates vary in practice, for example based on time-of-use, the interpretation of the demand charge results is simplified through a constant $R_{\text{EC}}$. Our proposed EMPC approach is generalizable to variable energy rates. 

The PV panel tilt is $5^{\circ}$, and the entire PV array faces southwest at an azimuth angle of $219^{\circ}$ from north. PV output is obtained from the System Advisor Model (SAM) \cite{Blair2018} for the year 2019. The solar resource is estimated using a $30$~minute resolution dataset from the National Solar Radiation Database (NSRDB) and interpolated to obtain a time series with $15$~min resolution. 

Real $15$~min resolution load data provided by the Port of San Diego for 2019 is utilized. On most days, the peak load occurs at night, as a large share of the load is from lighting. There is no significant difference in load profiles between weekends and weekdays.

MG optimization case studies are carried out using CVX, a package for solving convex programs in the MATLAB environment~\cite{cvx_2}. The simulation time step ($\Delta T$) was chosen as $15$~min, and the initial SOC value \textcolor{black}{$x(0)$} for January 1 is chosen as $0.5$. Finally, perfect forecasts are assumed for load and PV generation.

\subsection{Case setup}\label{Case setup}

\textcolor{black}{In Section~\ref{results}, the proposed EMPC (noted as ``EMPC'') is compared with traditional EMPC (noted as ``Trad'') considering different receding manners, horizon lengths,  and peak tracking setups:}

\textcolor{black}{\subsubsection{Shrinking and rolling horizons}\label{ShrinkingAndRolling}
For the shrinking horizon manner, the horizon length shrinks one time point each MPC step and gets restarted daily, while for the rolling horizon manner, the horizon length remains constant, with the forecast updated each MPC step for both cases. Each receding manner implements the~$50$\% SOC low threshold and the terminal SOC constraint differently, which leads to distinct BESS flexibility. Details of SOC constraint implementation for Trad and EMPC with shrinking and rolling horizons are explained separately.}

\textcolor{black}{First, the~$50$\% SOC low threshold is set at the end of the current day. For the shrinking horizon manner, it is at the end of the prediction horizon and the end of the reference horizon for Trad~\eqref{FinalSOC} and EMPC~\eqref{RefFinalSOC}, respectively. On the other hand, for the rolling horizon manner, the $50$\% SOC low threshold constraint stays at the same time point when the rolling horizon moves forward in time. Then, the $50$\% SOC low threshold gets closer to the beginning of the prediction horizon and the reference horizon for Trad and EMPC, respectively.}

\textcolor{black}{Secondly, as EMPC has two stages (the first reference stage that includes placing the $50$\% SOC low threshold, and the second MPC stage that includes placing the terminal SOC constraint), the shrinking and rolling horizon manners further impact the BESS performance. For EMPC to track the reference trajectory, following the $50$\% SOC low threshold placed in the first reference stage, a terminal SOC constraint~\eqref{terminalconstraint2} is set at the end of the prediction horizon during the MPC stage. For the shrinking horizon manner, the terminal SOC constraint overlaps with the $50$\% SOC low threshold~\eqref{RefFinalSOC} when the reference horizon and the prediction horizon have the same length. On the other hand, with rolling horizon, the terminal SOC constraint~\eqref{terminalconstraint2} is placed at the end of the prediction horizon, which extends beyond the time point where the $50$\% SOC low threshold is placed.}

\textcolor{black}{In summary, for EMPC with shrinking horizon, the~$50$\% SOC low threshold is enforced at the end of the prediction horizon during the MPC stage for all shrinking horizon cases (except for the case with $48$~h reference horizon and $24$~h prediction horizon), while for EMPC with rolling horizon, more BESS discharging can occur before the terminal SOC constraint~\eqref{terminalconstraint2} becomes active.} 

\textcolor{black}{\subsubsection{$24$ and $48$~h reference and prediction horizons}
For both receding manners, the results for both Trad and EMPC strategies are presented considering $24$ and $48$~h prediction horizons. In addition, the EMPC strategy considers $24$ and $48$~h reference horizons when the prediction horizon is $24$~h, and it considers a $48$~h reference horizon when the prediction horizon is $48$~h.}

\textcolor{black}{\subsubsection{Peak tracking}
For both receding manners, both Trad and EMPC for all horizon lengths are constructed with (``WT'') and without (``NT'') peak demand tracking. }

\textcolor{black}{\subsubsection{Results section outline}
In total, 20 cases are demonstrated in the next Section. Specifically, 10 cases with shrinking horizons are discussed in Section~\ref{SHsubsection} and Table~\ref{Table1}, and the other 10 cases with rolling horizons are discussed in Section~\ref{RHsubsection} and Table~\ref{Table2}. Furthermore, an ideal EMPC case with a $24$~h prediction horizon and a full-month reference trajectory with perfect forecast (noted as EMPC*) is shown in Tables~\ref{Table1} and~\ref{Table2}. EMPC* is an ideal case since information to generate a full-month reference trajectory is generally not available. However, EMPC* provides the best performance that can be achieved in this case study. The control actions obtained from EMPC* are cost-optimal since the reference horizon in EMPC* spans the entire month.}

\section{Results and Discussion}\label{results}

\subsection{Shrinking-horizon cases}\label{SHsubsection}

\subsubsection{Full month reference horizon}
 
\textcolor{black}{In this section, the ideal case, EMPC*, is compared with Trad and EMPC for shrinking horizon cases with the monthly BESS SOC analysis. While Trad with a shrinking horizon uses the $50$\% SOC low threshold as the terminal SOC constraint, EMPC with a shrinking horizon overlaps the $50$\% SOC low threshold with the terminal SOC constraint when the reference horizon and the prediction horizon have the same length. EMPC* places the $50$\% SOC low threshold at the end of the reference horizon, which corresponds to the full month. While the $50$\% SOC low threshold guarantees that the BESS always has capacity available for cost minimization the next day, which is outside the prediction horizon, the $50$\% SOC low threshold as a terminal SOC constraint limits the BESS energy allocation flexibility at the end of each prediction horizon.}

\begin{figure} 
\centering
 \includegraphics[trim={0 0 0 0},width=\columnwidth]{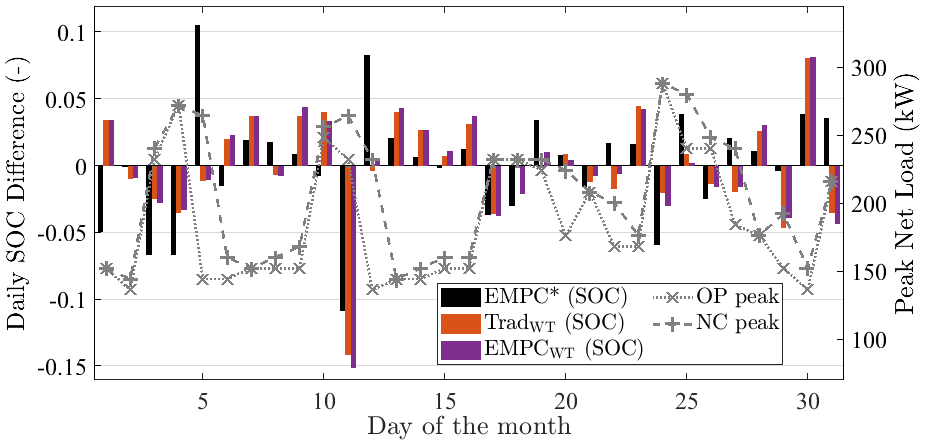}%
\caption{Daily SOC difference between final and initial SOC (colored bars) together with daily NC peak (dashed gray line with ``+'' markers) and OP peak net loads (dotted gray line with ``x'' markers) in March 2019 for EMPC* and the following shrinking horizon cases: $\text{Trad}_{\rm WT}$ with $T_{\text{MPC}}=48$~h, and $\text{EMPC}_{\rm WT}$ with $T_{\text{R}}=T_{\text{MPC}}=48$~h. Cases with $T_{\text{MPC}}=24$~h have daily SOC difference equals zero because SOC starts and ends at $50$\% every day.} 
\label{fig3:subfig1}
\vspace{-1em}
\end{figure}

The effect of the terminal SOC constraint on the demand charge reduction for all cases is further explained using the daily SOC difference between hours 00 and 24 in~Fig.~\ref{fig3:subfig1}. In March 2019, Fig.~\ref{fig3:subfig1} shows that larger daily SOC differences are observed for EMPC* compared to the other cases. Furthermore, large changes from negative SOC difference to positive SOC difference on consecutive days coincide with large drops in OP peak net load. For example, the SOC difference increases from $-0.06$ on day 4 to $0.1$ on day 5 as a result of the daily OP peak net load decreasing from $275$~kW to $175$~kW. Similar trends are observed on days 10, 11, and 24. With perfect knowledge of a full-month reference trajectory, EMPC* is able to adjust the SOC freely and allocate more energy on days with higher peak net load to reduce OP demand charges. 

\begin{table*}[]
\caption{Annual cost components for EMPC* and shrinking prediction horizon cases for the year 2019. The highest and the lowest annual costs are highlighted in red and bold, respectively.}
\begin{tabular}{  m{7em}  m{3em} | m{3em} m{3em} m{4em} m{4em} | m{3em} m{3em} m{4em} m{4em} m{4em} m{4em}  } 
Cases               & EMPC*    & $\text{Trad}_{\rm NT}$  & $\text{Trad}_{\rm WT}$ & $\text{EMPC}_{\rm NT}$   &  $\text{EMPC}_{\rm WT}$   & $\text{Trad}_{\rm NT}$ & $\text{Trad}_{\rm WT}$ & $\text{EMPC}_{\rm NT}$ &  $\text{EMPC}_{\rm WT}$   & $\text{EMPC}_{\rm NT}$ &  $\text{EMPC}_{\rm WT}$  \\
 \cline{1-12}
\textcolor{black}{Run time (s)}  & -    &0.7     &0.7     &1.4    &1.4   &1.0   &1.0   &1.5    &1.5   &3.1    &3.1 \\ 
  \cline{1-12}
$T_{\text{MPC}}$    & 24 h    & 24 h    & 24 h    & 24 h   & 24 h    & 48 h   & 48 h    & 24 h   & 24 h    & 48 h   & 48 h  \\
$T_{\text{R}}$      & 1 mo & -       & -          & 24 h   & 24 h    & -      & -       & 48 h   & 48 h    & 48 h   & 48 h  \\
 \cline{1-12}
NCDC (k\$)          & 43.0    & 51.7    & 39.5   & 58.9    & 39.5   & 46.3   & 38.2   & 53.9  & 38.2  & 46.5  & 38.2   \\
OPDC (k\$)          & \ph1.7  & 23.1    & 19.3   & 18.0    & 19.3    & 22.1   & 19.1  & 17.4  & 19.1  & 18.4  & 19.1   \\
Energy Cost (k\$)   & 14.4    & 14.4    & 14.4   & 14.4    & 14.4    & 14.4   & 14.4  & 14.4  & 14.4  & 14.4  & 14.4  \\
BESS loss (k\$)     & \ph5.5  & 10.0    & \ph4.9 & \ph3.3  & \ph4.9  & \ph9.0 & \ph5.0 & \ph4.9 & \ph5.0 & \ph4.1 & \ph5.0   \\
 \cline{1-12} 
Annual Cost (k\$)   &  64.6   & \textbf{\textcolor{purple}{99.2}}    & 78.1   & 94.5  & 78.1    & 91.8  & \textbf{76.7} & 90.6  & \textbf{76.7}  & 83.3 & \textbf{76.7}    \\
\end{tabular}
\label{Table1}
\end{table*}

\subsubsection{24-hour reference and prediction horizons}\label{24h SH subsubsection}
In this section EMPC is compared with Trad under the simplest settings- $24$~h, NT. $\text{EMPC}_{\rm NT}$ outperforms the benchmark case, $\text{Trad}_{\rm NT}$, with a lower annual cost (Table~\ref{Table1}). Even though the reference trajectory in $\text{EMPC}_{\rm NT}$ does not track the peak, in \eqref{yNC_OP_initial} EMPC includes information on the highest peaks up to time $t$ in the auxiliary variables $y_{\rm NC}$ and $y_{\rm OP}$. Thus, $\text{EMPC}_{\rm NT}$ has the peak information from a peak tracking method embedded within the formulation. When the highest monthly peak observed till time $t$ is higher than the predicted demand, compared to $\text{Trad}_{\rm NT}$, $\text{EMPC}_{\rm NT}$ dispatches the BESS less aggressively for demand charge reduction, reducing the BESS loss cost.

\begin{figure}[htp]
\centering 
\includegraphics[width=\columnwidth]{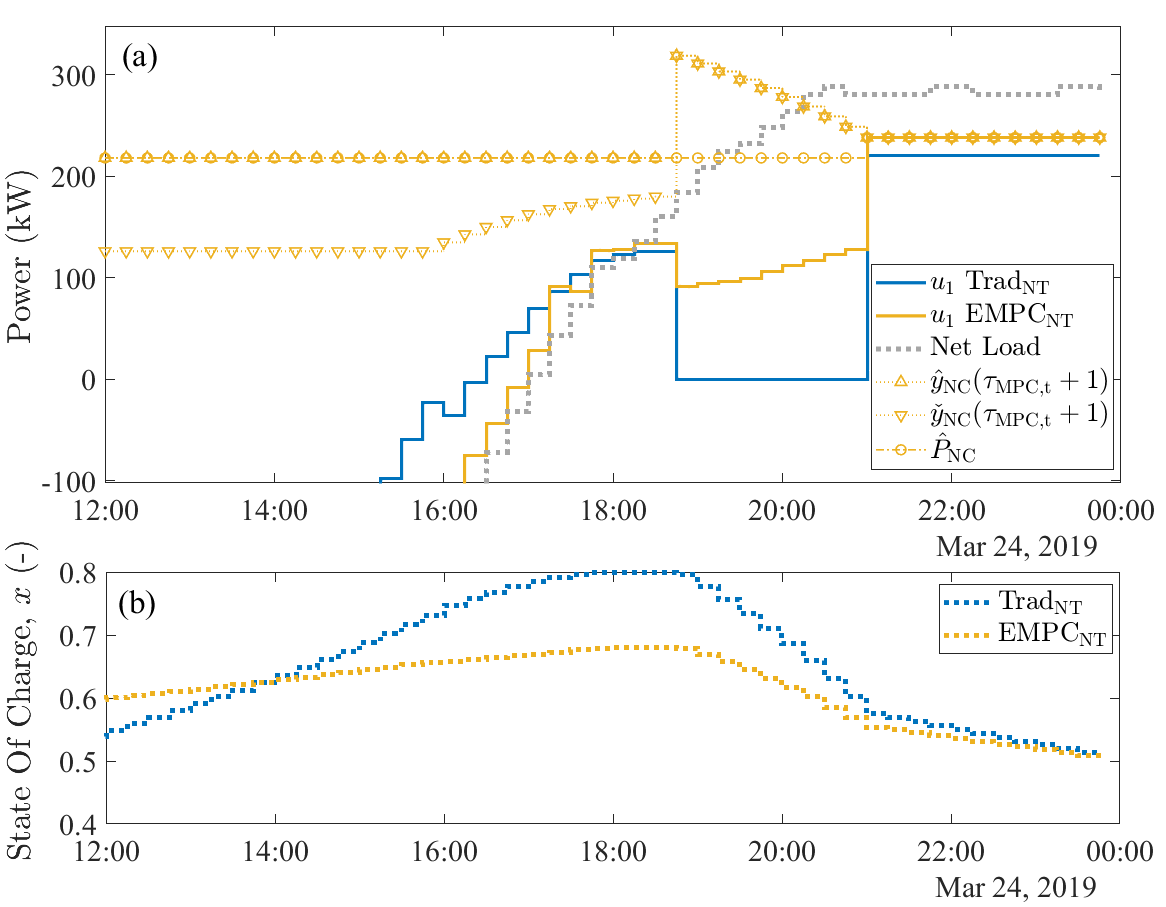}
\caption{March 24 timeseries of $\text{Trad}_{\rm NT}$ with $T_{\text{MPC}}=24$~h, and $\text{EMPC}_{\rm NT}$ with $T_{\text{MPC}}=T_{\text{R}}=24$~h shrinking horizon cases for: (a) demand ($u_1$, solid lines) together with $\hat{y}_{\rm NC}(\tau_{\text{MPC},t}+1)$, $\check{y}_{\rm NC}(\tau_{\text{MPC},t}+1)$, and $\hat{P}_{\text{NC}}(t)$ for $\text{EMPC}_{\rm NT}$; and (b) SOC ($x$, dotted lines). The dotted gray line in (a) represents the net load, $L_{f}-\text{PV}_{f}$. The net load before 16:00~h is less than -100~kW due to solar over-generation.}
\label{FigureShrinked_24h}
\end{figure}

For example, from $18$:$30$~h to $21$:$00$~h on March $24$, 2019 (Fig.~\ref{FigureShrinked_24h}), $\text{Trad}_{\rm NT}$ (blue line) dispatches the BESS aggressively for OP demand reduction, while $\text{EMPC}_{\rm NT}$ (yellow line) allows more OP demand while remaining below the previous highest OP demand (Fig.~\ref{FigureShrinked_24h}a). Compared to $\text{Trad}_{\rm NT}$, $\text{EMPC}_{\rm NT}$ controls the BESS less aggressively (Fig.~\ref{FigureShrinked_24h}b), maintaining the SOC close to 50\%, resulting in lower BESS losses. 

On the other hand, reduced BESS charging in EMPC causes issues when a high net load exists close to the end of the day. Compared to $\text{Trad}_{\rm NT}$, $\text{EMPC}_{\rm NT}$ has less BESS energy left that can be dispatched to reduce the NC peak. This results in a higher annual NCDC as shown in Table~\ref{Table1} (NCDC$=\text{k\$}58.9$ vs. k\$51.7). 
For example, in Fig.~\ref{FigureShrinked_24h}b the low SOC at 21:00~h and the terminal SOC constraint causes $\text{EMPC}_{\rm NT}$ to reach a NC peak of $238$~kW compared to the $\text{Trad}_{\rm NT}$ peak of $221$~kW. This higher NC demand close to the end of the day is observed in all EMPC cases throughout the year because the Port of San Diego consists of mostly unconditioned warehouses and outdoor lighting that trigger a higher net load after 21:00~h. 

Adding peak tracking to both control strategies, $\text{Trad}_{\rm WT}$ and $\text{EMPC}_{\rm WT}$ yield the same annual cost of k\$78.1 (Table~\ref{Table1}). 

\subsubsection{Longer reference and prediction horizons}\label{48h SH subsubsection} Similar trends are observed for the $48$~h horizons: without peak tracking, $\text{EMPC}_{\rm NT}$ with $T_{\text{R}}=T_{\text{MPC}}=48$~h outperforms $\text{Trad}_{\rm NT}$ with $T_{\text{MPC}}=48$~h by k\$8.5 ($=\text{k\$}91.8-\text{k}\$83.3$); and with peak tracking, $\text{EMPC}_{\rm WT}$ with $T_{\text{R}}=T_{\text{MPC}}=48$~h and $\text{Trad}_{\rm WT}$ with $T_{\text{MPC}}=48$~h have the same annual cost of k\$76.7.
 
The higher NCDC observed in $\text{EMPC}_{\rm NT}$ compared to $\text{Trad}_{\rm NT}$ in the previous section due to the terminal SOC constraint and reduced BESS charging can be mitigated by expanding both the prediction horizon and the reference horizon. Expanding the reference horizon from $24$~h to $48$~h reduces the annual NCDC from k\$$58.9$ to k\$$53.5$ for $\text{EMPC}_{\rm NT}$, and from k\$$39.5$ to k\$$38.2$ for $\text{EMPC}_{\rm WT}$. By further expanding the prediction horizon from $24$~h to $48$~h, the annual $\text{EMPC}_{\rm NT}$ NCDC is reduced from k\$$53.9$ to k\$$46.5$. As no terminal SOC constraint is set at the end of the first day, the BESS has more flexibility to allocate its energy for demand charge reduction. The benefit of prediction horizon expansion is also observed for Trad. Yet higher NCDC and lower BESS loss in $\text{EMPC}_{\rm NT}$ compared to $\text{Trad}_{\rm NT}$ are also observed for the $48$~h prediction horizon with NCDC~$=\text{k\$}46.5$ versus k\$46.3 and BESS loss~$=\text{k\$}4.1$ versus k\$9.0. Besides expanding the horizon, implementing a rolling horizon instead of a shrinking horizon also allows relaxing the terminal SOC constraint, which is further discussed in Section~\ref{RHsubsection}.

All cases presented in Table~\ref{Table1} and Table~\ref{Table2} share the same energy cost (excluding BESS losses), as the charging timing of BESS does not affect the energy cost as the energy rate is constant throughout the day.

\subsection{Rolling-horizon cases}\label{RHsubsection}
\subsubsection{Rolling versus shrinking horizons}
Results for NT with rolling horizons in Table~\ref{Table2} show that EMPC outperforms Trad for both $24$~h and $48$~h horizons, 
\textcolor{black}{which is}
consistent with what was observed for the shrinking horizon (Section~\ref{SHsubsection}). On the other hand, for WT the annual costs between EMPC and Trad differ, which is not the same pattern observed for shrinking horizons. The differences between the shrinking and rolling horizon cases are caused by the \textcolor{black}{difference in the placement of $50$\% SOC low threshold time point within the reference horizon (Section ~\ref{ShrinkingAndRolling}).}
\textcolor{black}{As the end of the prediction horizon shifts away from the $50$\% SOC low threshold of the reference horizon, more BESS discharging can occur before the terminal SOC constraint~\eqref{terminalconstraint2} becomes active. That is, the BESS is discharged deeper after the $50$\% SOC low threshold placement to reduce electricity costs, leading to low SOC at the end of the reference horizon, which is then assigned as the terminal SOC constraint by~\eqref{terminalconstraint2} in the prediction horizon. In this case, the reference SOC value used as the terminal SOC constraint in EMPC is close to the minimal SOC limit, $\underline{x}$.}

\subsubsection{24-hour reference and prediction horizons}\label{24h RH subsubsection}
\begin{figure}[t]
\centering 
\includegraphics[width=\columnwidth]{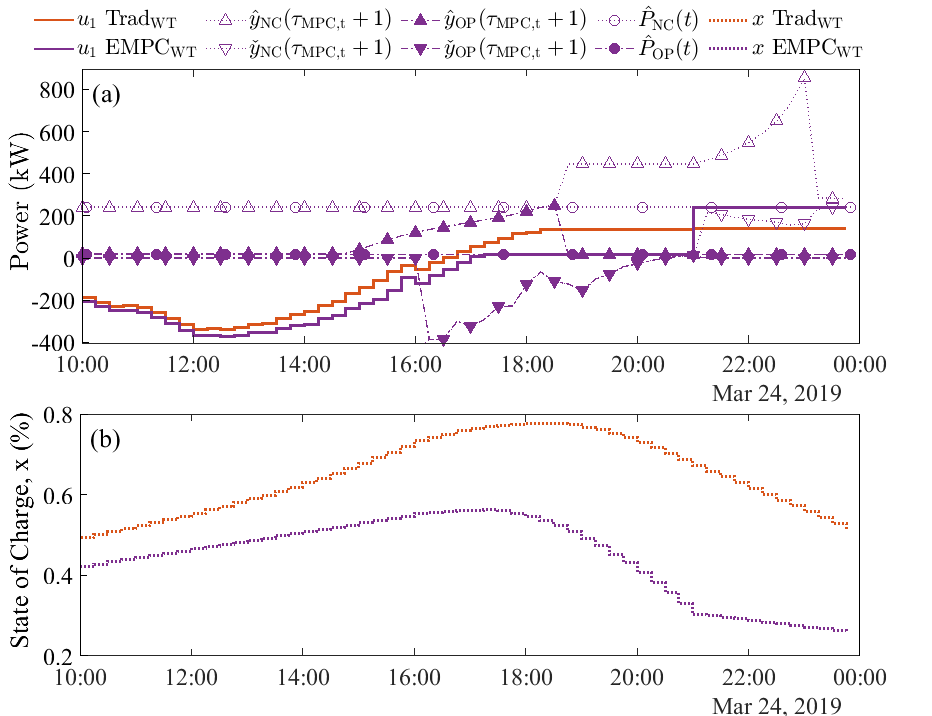}
\caption{\textcolor{black}{March 24 timeseries of (a) $u_1$, and (b) $x$, together with $\hat{y}_{\rm NC}(\tau_{\text{MPC},t}+1)$, $\check{y}_{\rm NC}(\tau_{\text{MPC},t}+1)$, $\hat{y}_{\rm OP}(\tau_{\text{MPC},t}+1)$, $\check{y}_{\rm OP}(\tau_{\text{MPC},t}+1)$, $\hat{P}_{\rm NC}(t)$, and $\hat{P}_{\rm OP}(t)$ for $\text{Trad}_{\rm WT}$ with $T_{\text{MPC}}=24$~h and $\text{EMPC}_{\rm WT}$ with $T_{\text{R}}=T_{\text{MPC}}=24$~h. All results are for rolling horizon cases.}}
\label{figRH1}
\end{figure}

For the $24$~h horizons, $\text{EMPC}_{\rm WT}$ has a higher annual cost than $\text{Trad}_{\rm WT}$ (k\$$91.5$ versus k\$$77.4$ in Table~\ref{Table2}) because of excessive BESS discharges during OP hours that generate larger NC peak demands at night in $\text{EMPC}_{\rm WT}$ compared to $\text{Trad}_{\rm WT}$ (NCDC$=\text{k\$}66.0$ versus k\$$39.4$ in Table~\ref{Table2}). Indeed, the difference in NCDC corresponds to the largest share of the annual cost for $\text{EMPC}_{\rm WT}$ and negates any reduction achieved in OPDC and BESS losses compared to $\text{Trad}_{\rm WT}$. Figure~\ref{figRH1} shows the March 24 $u_1$ timeseries as an example, where $\text{EMPC}_{\rm WT}$ shows a lower OP peak demand but a larger NC peak compared to $\text{Trad}_{\rm WT}$ (OP peak: $18$~kW versus $134$~kW, NC peak: $242$~kW versus $139$~kW, respectively). Figure~\ref{figRH1} also shows the difference in SOC between $\text{EMPC}_{\rm WT}$ and $\text{Trad}_{\rm WT}$, where the former reaches $30$\% SOC at 21:00~h while SOC for $\text{Trad}_{\rm WT}$ is $67$\%.

\begin{table*}[]
\centering
\caption{Annual cost components for EMPC* and rolling prediction horizon cases for the year 2019. The highest and the lowest annual costs are highlighted in red and bold, respectively.}
\begin{tabular}{  m{7em} m{3em} | m{3em} m{3em} m{4em}  m{4em} | m{3em}  m{3em}  m{4em}   m{4em}  m{4em}  m{4em}  } 
Cases     & $\text{EMPC*}$         & $\text{Trad}_{\rm NT}$ & $\text{Trad}_{\rm WT}$  & $\text{EMPC}_{\rm NT}$   & $\text{EMPC}_{\rm WT}$  & $\text{Trad}_{\rm NT}$ & $\text{Trad}_{\rm WT}$ & $\text{EMPC}_{\rm NT}$  & $\text{EMPC}_{\rm WT}$  & $\text{EMPC}_{\rm NT}$  & $\text{EMPC}_{\rm WT}$  \\
 \cline{1-12}
 \textcolor{black}{Run time (s)}            &-    &0.8     &0.8     &1.9    &1.9   &1.1   &1.1   &2.2    &2.2   &3.9    &3.9 \\ 
  \cline{1-12}
$T_{\text{MPC}}$    & 24 h    & 24 h   & 24 h   & 24 h   & 24 h   & 48 h   & 48 h   & 24 h   & 24 h   & 48 h   & 48 h  \\
$T_{\text{R}}$      & 1 mo & -      & -      & 24 h   & 24 h   & -      & -      & 48 h   & 48 h   & 48 h   & 48 h  \\
 \cline{1-12}
NCDC (k\$)          & 43.0    & 53.1   & 39.4   & 66.2   & 66.0   & 49.3   & 39.6   & 47.0   & 39.5   & 41.6   & 40.2  \\
OPDC (k\$)          & 1.7     & 21.9   & 18.4   & \ph8.9 & \ph7.5 & 20.3   & 17.5   & 18.8   & 17.5   & 18.9   & 15.3  \\
Energy Cost (k\$)       & 14.4    & 14.4   & 14.4   & 14.4   & 14.4   & 14.4   & 14.4   & 14.4   & 14.4   & 14.4   & 14.4  \\
BESS loss (k\$)     & \ph5.5  & \ph9.9 & \ph5.1 & \ph3.4 & \ph3.5 & \ph9.0 & \ph5.0 & \ph4.0 & \ph5.0 & \ph4.5 & \ph5.1  \\
 \cline{1-12} 
Annual Cost (k\$)   &64.6 & \textbf{\textcolor{purple}{99.3}}  & 77.4   & 92.9   & 91.5  & 92.9  & 76.5  & 84.2  & 76.5  & 79.5  & \textbf{74.9}  \\
\end{tabular}
\label{Table2}
\vspace{-2em}
\end{table*}
  
The aggressive BESS discharge during OP hours for $\text{EMPC}_{\rm WT}$ are a consequence of the reference trajectory suggesting OP and NC peak demand increments within the prediction horizon. The reference trajectory suggests an NC (OP) peak demand increment within the prediction horizon when $\hat{y}_{\rm NC}(\tau_{\text{MPC},t}+1) > \hat{P}_{\text{NC}}(t) > \check{y}_{\rm NC}(\tau_{\text{MPC},t}+1)$ ($\hat{y}_{\rm OP}(\tau_{\text{MPC},t}+1) > \hat{P}_{\rm OP}(t) > \check{y}_{\rm OP}(\tau_{\text{MPC},t}+1)$). Figure~\ref{figRH1} shows that the reference trajectory starts suggesting increments at 14:45~h for OP hours and at 18:30~h for NC hours.  

The reference trajectory proposes increments in OP and NC peaks within the prediction horizon because of the proximity of the $50$\% SOC low threshold to time $t$ in the reference horizon. Indeed, a low SOC as an initial condition for the reference trajectory generates aggressive reference BESS charging at the beginning of the reference horizon, and also aggressive discharging towards the end of the reference horizon. The EMPC algorithm with increments in NC and OP peaks within the prediction horizon as a reference and a low terminal SOC state will aggressively discharge the BESS to avoid peak demand increments within the prediction horizon (and minimize the terminal cost $V_f$).  

\subsubsection{Longer reference and prediction horizons}\label{48h RH subsubsection}
\begin{figure}[t]
\centering 
\includegraphics[width=\columnwidth]{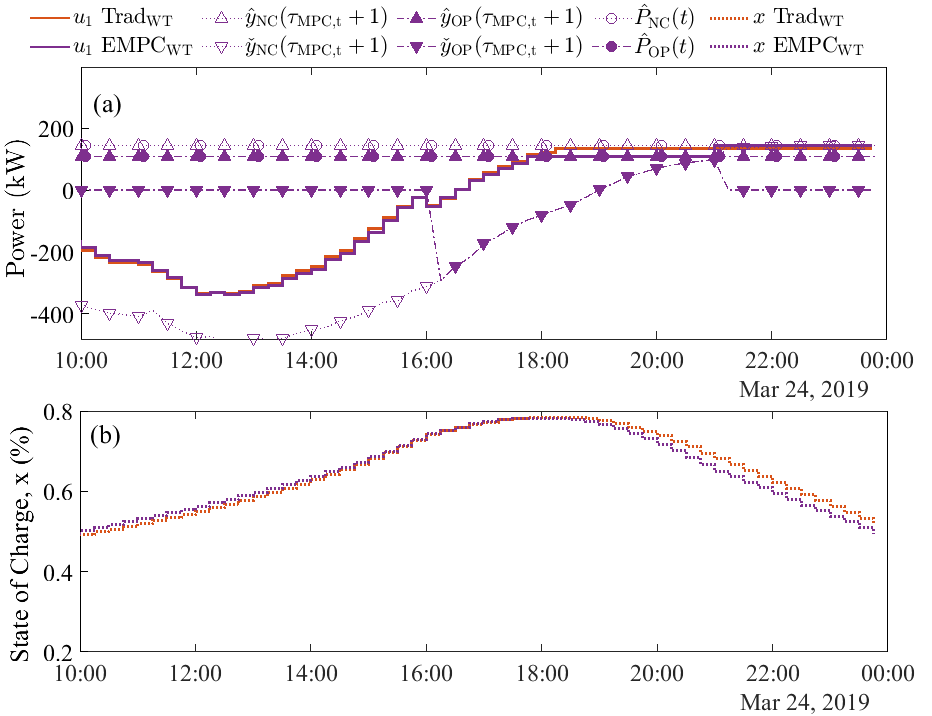}
\caption{\textcolor{black}{March 24 timeseries of (a) $u_1$, and (b) $x$, together with $\hat{y}_{\rm NC}(\tau_{\text{MPC},t}+1)$, $\check{y}_{\rm NC}(\tau_{\text{MPC},t}+1)$, $\hat{y}_{\rm OP}(\tau_{\text{MPC},t}+1)$, $\check{y}_{\rm OP}(\tau_{\text{MPC},t}+1)$, $\hat{P}_{\rm NC}(t)$, and $\hat{P}_{\rm OP}(t)$ for $\text{Trad}_{\rm WT}$ with $T_{\text{MPC}}=48$~h and $\text{EMPC}_{\rm WT}$ with $T_{\text{R}}=T_{\text{MPC}}=48$~h. All results are for rolling horizon cases.}}
\label{figRH2}
\end{figure}

The explanation provided for the $24$~h horizons also applies to $\text{EMPC}_{\rm WT}$ with $T_{\text{R}}=T_{\text{MPC}}=48$~h, but here a 2\% reduction in the annual cost is achieved with respect to $\text{Trad}_{\rm WT}$ with $T_{\text{MPC}}=48$~h. The main cost reduction in $\text{EMPC}_{\rm WT}$ comes from a lower OPDC ($12.6$\% lower than $\text{Trad}_{\rm WT}$), while NCDC only increases $1.5$\%, as shown in Table~\ref{Table2}. The annual cost reduction for the $48$~h horizon is a result of BESS discharge during OP hours in $\text{EMPC}_{\rm WT}$ that is more moderated than the case of EMPC with $24$~h reference horizon but still more aggressive than the $\text{Trad}_{\rm WT}$ with $T_{\text{MPC}}=48$~h. As an example, Fig.~\ref{figRH2} shows $\text{EMPC}_{\rm WT}$ with an OP peak of $108$~kW (NC peak of $143$~kW), while an OP peak of $134$~kW (NC peak of $134$~kW) is obtained with $\text{Trad}_{\rm WT}$. The $\text{EMPC}_{\rm WT}$ improvement explained above is not observed with $T_{\text{R}}=48$~h and $T_{\text{MPC}}=24$~h because in this case the prediction horizon ends before the $50$\% SOC low threshold is placed in the reference horizon.

$\text{EMPC}_{\rm WT}$ with $T_{\text{R}}=T_{\text{MPC}}=48$~h implements a more moderated BESS discharge than $\text{EMPC}_{\rm WT}$ with $T_{\text{R}}=T_{\text{MPC}}=24$~h because of the longer reference and prediction horizons. The $48$~h reference trajectory avoids suggesting peak demand increments within the EMPC prediction horizon because the $50$\% threshold is at least $24$~h ahead of the reference horizon initial time point at every MPC step. No NC (OP) peak increments within the prediction horizon are suggested by the reference trajectory when $\hat{y}_{\rm NC}(\tau_{\text{MPC},t}+1) = \hat{P}_{\rm NC}(t)$ ($\hat{y}_{\rm OP}(\tau_{\text{MPC},t}+1) = \hat{P}_{\rm OP}(t)$). The $48$~h prediction horizon includes two OP time periods with a 19~h gap in between that reduces BESS discharging in the first OP period to keep enough energy for the second OP period. Thus, the $48$~h prediction horizon prevents the BESS from reaching low SOC levels at the end of the first OP time period. As an example, Fig.~\ref{figRH2} shows $\text{EMPC}_{\rm WT}$ with $T_{\text{R}}=T_{\text{MPC}}=48$~h at a SOC of $65$\% at 21:00~h, while at the same time a $68$\% SOC is reached by $\text{Trad}_{\rm WT}$ with $T_{\text{MPC}}=48$~h.

\subsection{Run time comparison and scalability} \label{RunTimeScalability}

\textcolor{black}{Table~\ref{Table1} and Table~\ref{Table2}} shows that
the average run time of a single simulation step for EMPC is $1.4$~s ($3.1$~s) for $24$~h ($48$~h) shrinking reference and prediction horizons, whereas the average run time for Trad is $0.7$~s ($1.0$~s) for the same horizon configurations. For rolling horizons, the average run time is $1.9$~s ($3.9$~s) for EMPC and $0.8$~s ($1.1$~s) for Trad with $24$~h ($48$~h) reference and prediction horizons. \textcolor{black}{Note that a decrease of prediction horizon length, while reducing computational cost for more efficient real-time implementation means that less future information is available for MPC and thus may lead to suboptimal economic performance.}

\textcolor{black}{The proposed method in Section~\ref{EMPCsubsection} requires solving two nominal EMPC problems based on convex optimization (one for the reference trajectory, and another for real-time dispatch) at each simulation step for real-time execution. Convex optimization-based EMPC for systems such as the one proposed in this method has a low computational cost, and a variety of tools (like CVX, YALMIP) are available to get accurate solutions efficiently~\cite{boyd2004convex}. The case study in Section~\ref{CaseStudy} considers one PV system and a BESS in a grid connected microgrid (MG). However, the proposed method can be extended to multiple dispatchable units within a MG or multiple MGs. The only requirement for scaling up the model is the inclusion of additional constraints: equality constraints to capture the state of charge dynamics of BESS systems, and inequality constraints to account for the power ratings of BESS and PV systems. For example, two PV and two BESS systems over a $24$~h prediction horizon (with $T_{\text{MPC}}=96$ discrete time steps and $\Delta T=15$~min) entails adding just $T_{\text{MPC}}$ optimization variables (for the second BESS dispatch power) and $3T_{\text{MPC}}$ constraints (for the second BESS maximum and minimum charging/discharging power limits and SOC update) for each MPC step, which makes the problem scale linearly. Convex optimization problems involving 1,000 - 10,000s variables and constraints can be reliably solved by standard solvers on a single machine~\cite{boyd2004convex, Boyd}. The low run-time of a single simulation step (corresponding to 15 min in real life) varying between $1.4-3.9$~s for the different prediction horizons and shrinking/rolling horizon manners in the case study, also serves as a practical reinforcement of the scalability of the proposed method.}

\subsection{\textcolor{black}{Recursive feasibility and stability comparison with~\cite{Risbeck2020}}}\label{compare_risbeck}

\textcolor{black}{In~\cite[Eq. (2)]{Risbeck2020}, the MPC is parametrized by only those initial states that are able to give a feasible control input sequence over the prediction horizon, which enforces recursive feasibility by construction. In our work, as evidenced by~\eqref{terminalconstraint2}, the terminal state set for the MPC is equal to the value of the predicted reference state at the time point corresponding to the terminal time point of the MPC stage. For our case studies, the entire state set (SOC range) from 0.2 to 0.8 can be traversed in about $9$ time steps. The prediction horizon varies between $24$ to $48$~h, which makes the time steps within an MPC prediction horizon vary between $96-192$ steps. Thus, recursive feasibility is always attained as long as the prediction horizon is more than $9$ time steps, which like~\cite{Risbeck2020} is enforced by construction of the optimal control problem in our work.}

\textcolor{black}{\cite{Risbeck2020} established that the region of attraction\footnote{\textcolor{black}{The region of attraction is the set of initial states from which asymptotic stability of the closed-loop solution of the MPC is achieved with respect to the known reference trajectory.}} under their control law is nothing other than the feasible initial state set under assumptions of strict dissipativity, stability of the terminal region and a-priori knowledge of the reference trajectory for the entire window for which the demand charge is assessed (i.e., a month). For our work, for both the shrinking and rolling horizon cases, stability (the way it is defined in a classical sense in~\cite{Risbeck2020}) with respect to the reference trajectory is an undefined problem and has no useful result, as the reference trajectory changes for each time step for the rolling horizon, and each $24$ or $48$ h period for the shrinking horizon cases.}

\section{Conclusion and Future Work} \label{conclusion}

The proposed EMPC is a more practical implementation of the original formulation developed by \cite{Risbeck2020} for optimal BESS dispatch in a microgrid with variable renewable energy resources to minimize demand charges. The formulation from \cite{Risbeck2020} requires one month forecasts of renewable energy and load to generate the reference trajectory which is generally not available. On the other hand, we test 24~h and 48~h horizons for which forecasts are available in practice. Annual simulations show that even using a reference trajectory that is shorter than the demand charge assessment window, which is updated at every MPC step, the proposed EMPC algorithm still provides a total cost that is equal to (for shrinking horizons) or better than (for rolling horizons) the traditional EMPC when the monthly peak demands are tracked. The only exception is the rolling horizon case with a $24$~h reference horizon. For future implementations, the proposed EMPC with peak demand tracking and 48~h prediction and reference rolling horizons is preferred as it showed superior performance with respect to the traditional EMPC. Finally, although we make no claim about the generality of these economic benefits where the proposed EMPC performs no worse than the traditional EMPC, the results show that microgrid operators can reduce operating electricity costs using the proposed EMPC. 

The authors in~\cite{Risbeck2020} proved that the reference trajectory provides an upper bound on the asymptotic operating electricity costs when considering infinite time horizon EMPC problems with rolling horizon time windows. The results presented in this paper show that these results appear to carry over to finite time horizon EMPC problems as well for both shrinking and rolling horizons. Future work will focus on theoretical analysis of the asymptotic cost performance of the proposed EMPC. Further work may also consider generating a reference trajectory by solving an optimization problem with a different objective function, such as $\rm CO_2$ emission reduction. Additionally, incorporating errors into the demand and PV generation forecasts is needed to evaluate the proposed algorithm in real-world implementations.

\section*{Acknowledgments}

We acknowledge funding from the California Energy Commission under EPC-17-049. CC acknowledges financial support from CONICYT PFCHA/DOCTORADO BECAS CHILE/2017.
\vspace{-1em}
\bibliographystyle{IEEEtran}
\bibliography{refs}

\end{document}